\documentclass[a4paper,11pt]{article}
\usepackage[affil-sl,auth-sc]{authblk}
\usepackage{amssymb,amsmath,amsbsy}
\usepackage{mathbbol}
\usepackage{bm}
\usepackage{appendix}
\usepackage[linktocpage=true]{hyperref}
\usepackage[top=1.3in, bottom=1.3in,left=0.9in,right=0.9in]{geometry}   
\usepackage{graphicx}
\usepackage{cite}
\usepackage{float}
\usepackage{caption}
\usepackage{wasysym} \usepackage{float} \usepackage{slashed}
\usepackage{caption}
\usepackage{multirow}
\usepackage[nottoc]{tocbibind} 

\usepackage{color}

\newcommand{\nc}{\newcommand}

\newcommand{\bra}[1]{\langle #1|} \newcommand{\ket}[1]{|#1\rangle}
 \newcommand{\ie}{{\it
    i.e., }}  
 \newcommand{\bea}{\begin{eqnarray}}
\newcommand{\eea}{\end{eqnarray}} 
 
 \nc{\bce}{\begin{center}}
\nc{\ece}{\end{center}} \nc{\be }{\begin{equation}} \nc{\ee
}{\end{equation}} \nc{\btb}{\begin{tabular}} \nc{\etb}{\end{tabular}}
\nc{\f}{\frac} \nc{\eps}{\varepsilon} \nc{\vp}{\varphi}
\def\lcal{{\cal L}}  \nc{\tvp}{\widetilde{\varphi}}
 \nc{\vpj }{\mbox{${\vp^\dag
      i\,\raisebox{2mm}{\boldmath ${}^\leftrightarrow$}\hspace{-4mm}
      D_\mu\,\vp}$}} \nc{\vpjt}{\mbox{${\vp^\dag
      i\,\raisebox{2mm}{\boldmath ${}^\leftrightarrow$}\hspace{-4mm}
      D_\mu^{\,I}\,\vp}$}}
\newcommand{\hgg}{\mbox{$h\to \gamma\gamma$~}}

\def\eq#1{eq.~(\ref{#1})} \def\Eq#1{Eq.~(\ref{#1})}

\def\eqs#1#2{eqs.~(\ref{#1}) and (\ref{#2})}
\def\eqss#1#2#3{eqs.~(\ref{#1}), (\ref{#2}) and (\ref{#3})}

 \def\Ref#1{ref.~\cite{#1}}
 \def\Refs#1{refs.~\cite{#1}}

\let\originalleft\left \let\originalright\right
\renewcommand{\left}{\mathopen{}\mathclose\bgroup\originalleft}
\renewcommand{\right}{\aftergroup\egroup\originalright}

\numberwithin{equation}{section}

\begin{document}
  

\title{\bf The decay $h\to \gamma\gamma$ in the Standard-Model
  \\ Effective Field Theory}

\author{A. Dedes$^{1}$\footnote{email: {\tt adedes@cc.uoi.gr}},
  M. Paraskevas$^{1}$\footnote{email: {\tt mparask@grads.uoi.gr}},
  J. Rosiek$^{2}$\footnote{email: {\tt janusz.rosiek@fuw.edu.pl}},
  K. Suxho$^{1}$\footnote{email: {\tt csoutzio@cc.uoi.gr}} ~and~
  L. Trifyllis$^{1}$\footnote{email: {\tt ltrifyl@cc.uoi.gr}}}
\affil{\small $^{1}$Department of Physics, Division of Theoretical
  Physics, \\ University of Ioannina, GR 45110, Greece}
%
%
\affil{\small $^{2}$Faculty of Physics Department, University of
  Warsaw, Pasteura 5, 02-093 Warsaw, Poland}

\date{\today}

\maketitle
\thispagestyle{empty}

\begin{abstract}
Assuming that new physics effects are parametrized by the 
Standard-Model Effective Field Theory (SMEFT) 
written in a complete basis of up
to dimension-6 operators, we calculate the CP-conserving one-loop
amplitude for the decay $h\to \gamma\gamma$ in general $R_\xi$-gauges.
We employ a simple renormalisation scheme that is hybrid between
on-shell SM-like renormalised parameters and running
$\overline{\mathrm{MS}}$ Wilson coefficients.
The resulting amplitude is then finite, renormalisation scale
  invariant, independent of the gauge choice ($\xi$) and respects
SM Ward identities.
Remarkably, the $S$-matrix amplitude calculation resembles very
closely the one usually known from renormalisable theories and can be
automatised to a high degree.
We use this gauge invariant amplitude and recent LHC data to check
upon sensitivity {to} various Wilson coefficients entering from a more
complete theory at the matching energy scale.
We present a closed expression for the ratio $\mathcal{R}_{h\to
  \gamma\gamma}$, of the Beyond the SM versus the SM contributions as
appeared in LHC $h\to \gamma\gamma$ searches.  The most important
contributions arise at tree level from the operators $Q_{\varphi B},
Q_{\varphi W}, Q_{\varphi WB}$, and at one-loop level from the dipole
operators $Q_{uB},Q_{uW}$. Our calculation shows also that, for
operators that appear at tree level in SMEFT, one-loop corrections can
modify their contributions by less than 10\%.  Wilson coefficients
corresponding to these five operators are bounded from current LHC
$h\to \gamma\gamma$ data -- in some cases an order of magnitude
stronger than from other searches. 
Finally, we correct results that appeared previously in the literature.
\end{abstract}

\newpage 

\tableofcontents
\newpage 

\section{Introduction}
\label{sec:intro}

The discovery of the Higgs boson~\cite{Higgs:1964pj, Englert:1964et,
  Guralnik:1964eu} in year 2012 was made possible mainly because of
its decay into two photons~\cite{ATLAS:2012gk,CMS:2012gu}.  The
current outcome for this decay channel from LHC (Run-2) with
center-of-mass energy $\sqrt{s}=13\;\text{TeV}$, integrated luminosity
of $36.1\;\text{fb}^{-1}$ and Higgs boson mass, $M_h=125.09 \pm
0.24\;\text{GeV}$ is summarised as the ratio between the
{experimentally measured value (which may include contributions from
  new physics scenarios)} relative to the Standard-Model (SM)
predicted value~\cite{Ellis:1975ap,Shifman:1979eb}
  \begin{equation}
\mathcal{R}_{h\to \gamma\gamma} = \frac{\Gamma(\text{EXP},h\to
  \gamma\gamma)}{\Gamma(\text{SM},h\to \gamma\gamma)} \,.
 \label{Rhgg0}
\end{equation}
%
The most recent measurements are presented by
ATLAS~\cite{Aaboud:2018xdt} and CMS~\cite{Sirunyan:2018ouh}
experiments of LHC,
\begin{align}
\text{ATLAS:}\qquad \mathcal{R}_{h\to \gamma\gamma}&= 0.99^{ +
  0.15}_{ - 0.14}\,, \nonumber \\[1mm]
\text{CMS:}\qquad \mathcal{R}_{h\to \gamma\gamma}&= 1.18^{ + 0.17}_{ - 0.14}\,, 
\label{Rhgg}
\end{align}  
and are consistent with the SM prediction, with the error margin expected
to be reduced in the near future.

If we consider the SM as a complete theory of electroweak (EW) and
strong interactions up to the Planck scale, with no other scale
involved in between, then the decay amplitude $h\to \gamma\gamma$
arises purely from dimension $d\le 4$ (renormalisable) interactions.
In this case the amplitude is finite, calculable and, since all
relevant parameters are experimentally known, it is a certain
prediction of the SM. It is this prediction entering the denominator
in \eq{Rhgg0}.
If however, there is New Physics beyond the SM already at a scale
$\Lambda$ which is above, but not far from, the EW scale, say $\Lambda
\sim {\cal O}(1-10)~\mathrm{TeV}$, then its effects can be
parametrized by the presence of effective operators with dimension $d
> 4$ at scale $\Lambda$.  These operators together with various
parameters (or Wilson coefficients) will then run down to the EW scale
and feed the on-shell scattering $S$-matrix amplitude together with
$d\le 4$ interactions.

All dimension $d\le 6$ effective operators among SM particles that
obey the SM gauge symmetry have been classified in
refs.~\cite{Buchmuller:1985jz, Grzadkowski:2010es}.  The SM augmented
with these effective operators -- remnants of unknown heavy particles'
decoupling~\cite{Appelquist:1974tg} -- is called the SM Effective
Field Theory, or for a short SMEFT.  The quantization of SMEFT has
recently been undertaken in ref.~\cite{Dedes:2017zog} in linear
$R_\xi$-gauges with explicit proof of BRST symmetry and where all
relevant primitive interaction vertices have been collected.

Within SM, numerous calculations for the $h\to \gamma\gamma$ amplitude
exist.  The first calculation was performed in
ref.~\cite{Ellis:1975ap} in the limit of light Higgs mass ($M_{h}\ll
M_{W}$), using dimensional regularisation in the 't Hooft-Feynman
gauge.  Since then, there are other works completing this
calculation in linear and non-linear gauges~\cite{Ioffe:1976sd,
  Shifman:1979eb, Gavela:1981ri}, with different regularisation
schemes~\cite{Huang:2011yf, Shao:2011wx, Bursa:2011aa,
  Piccinini:2011az, Dedes:2012hf, Cherchiglia:2012zp, Donati:2013iya}.
To our knowledge the complete SM one-loop $h\to \gamma\gamma$
amplitude in linear $R_\xi$-gauges is performed in
ref.~\cite{Marciano:2011gm}.

In SMEFT\footnote{For a recent review see, \Ref{Brivio:2017vri} and
  for pedagogical lectures \Ref{Manohar:2018aog}.}  there is already a
number of papers that calculate the $h\to \gamma\gamma$
amplitude~\cite{Manohar:2006gz, Grojean:2013kd,Ghezzi:2015vva,Vryonidou:2018eyv}.\footnote{For earlier attempts see, \Refs{Perez:1992sa, Hernandez:1995vu}.}$^{,}$\footnote{Also, recently,
   the one-loop calculation for $h\to ZZ$ and $h\to
  Z\gamma$ decay in SMEFT has appeared in ref.\cite{Dawson:2018pyl}.}  The current, state of the art
calculation, has been presented by Hartmann and Trott in
refs.~\cite{Hartmann:2015aia,Hartmann:2015oia}.  The analysis was
carried out using the Background Field Method
(BFM)~\cite{Abbott:1980hw}\footnote{For a more recent  
approach on BFM-SMEFT see ref.\cite{Helset:2018fgq}.} 
consistent with minimal subtraction
renormalisation scheme ($\overline{\mathrm{MS}}$) and included all
relevant (CP-conserving) dimension $d\le 6$ operators in calculating
finite, non-log parts of the diagrams.
Our work here is complementary but incorporates some additional
features of importance:
\begin{itemize}
\item a simple calculational treatment in linear $R_\xi$-gauges based
  on Feynman rules of ref.~\cite{Dedes:2017zog},
\item an analytical proof of gauge invariance (independence on the
  gauge choice $\xi$-parameter(s)) of the $S$-matrix element,
\item a simple renormalisation framework which leads to a finite and
  renormalisation scale invariant amplitude,
\item a compact semi-analytical expression highlighting the effect of
  new operators in the ratio $\mathcal{R}_{h\to \gamma\gamma}$ and
  corresponding bounds on Wilson coefficients.
\end{itemize} 
There are quite a few papers addressing a global fit to the Higgs data
from LHC Run-1 and Run-2 in the SMEFT framework~\cite{Murphy:2017omb,
  Jana:2017hqg, Ellis:2018gqa}.  Our work provides a simple
semi-analytic one-loop formula for the ratio $\mathcal{R}_{h\to
  \gamma\gamma}$ in eq.~\eqref{Rhgg0} that can be used by these
(usually tree level) fits or by analogous experimental analysis at LHC
for Higgs boson searches.

Our paper is organised as follows. In section~\ref{sec:operators} we
list operators contributing to the decay $h\to \gamma\gamma$ in
SMEFT. Next, in section~\ref{sec:renormalization} we develop, in a
pedagogical fashion, the renormalisation scheme for calculating the
$h\to \gamma\gamma$ amplitude. In section~\ref{sec:amplitude} we give
analytical expressions for all types of SM and SMEFT contributions to
the $h\to \gamma\gamma$ amplitude and to the ratio $\mathcal{R}_{h\to
  \gamma\gamma}$.  Semi-analytical prediction for $\mathcal{R}_{h\to
  \gamma\gamma}$, depending on the running Wilson coefficients and
renormalisation scale $\mu$, are collected in
section~\ref{sec:results}, and supplied with a discussion on numerical
constraints of these coefficients. We conclude in
section~\ref{sec:conclusions}. Finally, in Appendix~\ref{app:pv} we
collect analytical expressions for the relevant one-loop self-energies
and, relevant to $h\to \gamma\gamma$, three-point one-loop corrections
in general $R_\xi$-gauges.

\section{Relevant Operators}
\label{sec:operators}

In EFT, an effect from the decoupling of heavy particles with masses
of order $\Lambda$ is captured by the running parameters of the low
energy theory influenced by higher dimensional operators added to SM
renormalisable Lagrangian $\lcal_{\text{SM}}^{(4)}$.  The full
effective Lagrangian we consider here can be expressed as,
\begin{equation}
\lcal = \lcal_{\text{SM}}^{(4)} + \sum_{X} C^{X} Q_X^{(6)}
+ \sum_{f} C^{'f} Q_f^{(6)} \,,
\label{Leff}
\end{equation}
where $Q_X^{(6)}$ denotes dimension-6 operators that do not involve
fermion fields, while $Q_f^{(6)}$ denotes operators that contain
fermion fields.  All Wilson coefficients should be rescaled by
$\Lambda^2$, for example $C^X \to C^X/\Lambda^2$. We shall restore
$1/\Lambda^2$ only in section~\ref{sec:amplitude} and thereafter. The
prime in $C^{' f}$, denotes a coefficient in flavour (``Warsaw'')
basis of ref.~\cite{Grzadkowski:2010es} while we use unprimed
coefficients in fermion mass basis defined in
ref.~\cite{Dedes:2017zog}.
\begin{table}[t] 
\centering
\renewcommand{\arraystretch}{1.5}
\btb{||c|c||c|c||c|c||} 
\hline \hline
\multicolumn{2}{||c||}{$X^3$} & 
\multicolumn{2}{|c||}{$\vp^6$~ and~ $\vp^4 D^2$} &
\multicolumn{2}{|c||}{$\psi^2\vp^3$}\\
\hline
$Q_W$ & $\eps^{IJK} W_\mu^{I\nu} W_\nu^{J\rho} W_\rho^{K\mu}$ & 
$Q_\vp$ & $(\vp^\dag\vp)^3$ &
$Q_{e\vp}$ & $(\vp^\dag \vp)(\bar l'_p e'_r \vp)$\\
 &  & 
$Q_{\vp\Box}$ & $(\vp^\dag \vp)\raisebox{ - .5mm}{$\Box$}(\vp^\dag \vp)$ &
$Q_{u\vp}$ & $(\vp^\dag \vp)(\bar q'_p u'_r \tvp)$\\
& & 
$Q_{\vp D}$ & $\left(\vp^\dag D^\mu\vp\right)^* \left(\vp^\dag D_\mu\vp\right)$ &
$Q_{d\vp}$ & $(\vp^\dag \vp)(\bar q'_p d'_r \vp)$\\
%
%
\hline \hline
\multicolumn{2}{||c||}{$X^2\vp^2$} &
\multicolumn{2}{|c||}{$\psi^2 X\vp$} &
\multicolumn{2}{|c||}{$\psi^2\vp^2 D$}\\ 
\hline
$Q_{\vp B}$ & $ \vp^\dag \vp\, B_{\mu\nu} B^{\mu\nu}$ & 
$Q_{eW}$ & $(\bar l'_p \sigma^{\mu\nu} e'_r) \tau^I \vp W_{\mu\nu}^I$ &
$Q_{\vp l}^{(3)}$ & $(\vpjt)(\bar l'_p \tau^I \gamma^\mu l'_r)$ 
\\
%
 $Q_{\vp W}$&  $\vp^\dag \vp\, W^I_{\mu\nu} W^{I\mu\nu}$& 
$Q_{eB}$ & $(\bar l'_p \sigma^{\mu\nu} e'_r) \vp B_{\mu\nu}$ & &
\\
$Q_{\vp WB}$ &  $ \vp^\dag \tau^I \vp\, W^I_{\mu\nu} B^{\mu\nu}$& 
$Q_{uW}$ & $(\bar q'_p \sigma^{\mu\nu} u'_r) \tau^I \tvp\, W_{\mu\nu}^I$ & &
\\
 &  &
$Q_{uB}$ & $(\bar q'_p \sigma^{\mu\nu} u'_r) \tvp\, B_{\mu\nu}$ & &
\\
 &  &
$Q_{dW}$ & $(\bar q'_p \sigma^{\mu\nu} d'_r) \tau^I \vp\, W_{\mu\nu}^I$ & &
\\
& & 
$Q_{dB}$ & $(\bar q'_p \sigma^{\mu\nu} d'_r) \vp\, B_{\mu\nu}$ & & \\
\hline \hline
\multicolumn{2}{||c||}{} & \multicolumn{2}{|c||}{$\psi^4$} & \multicolumn{2}{|c||}{}\\
\hline
\multicolumn{2}{||c||}{} & $Q_{ll}$ & $(\bar l'_p \gamma_\mu
l'_r)(\bar l'_s \gamma^\mu l'_t)$ & \multicolumn{2}{|c||}{} \\
\hline\hline
\etb
\caption{\sl A set of $d=6$ operators in Warsaw basis that contribute
  to the $h\to \gamma\gamma$ decay amplitude, directly or indirectly,
  in $R_\xi$-gauges. We consider only CP-conserving operators in our
  analysis.  The operator $Q_{\varphi}$ cancels out completely in the
  \hgg amplitude.  The operators $Q_{ll}$ and $Q_{\varphi l }^{(3)}$
  present themselves indirectly through the translation of the
  renormalised vacuum expectation value (vev) into the well measured
  Fermi coupling constant, {\it cf.}~\eq{eq:GF}.  The notation is the
  same as in refs.~\cite{Grzadkowski:2010es, Dedes:2017zog}.  For
  brevity we suppress fermion chiral indices $L,R$.}
   \label{tab:no4ferm}
\end{table}

The operators involved in the calculation of decay $h\to
\gamma\gamma$ are collected in Table~\ref{tab:no4ferm}.  They can
easily be identified when drawing the Feynman diagrams for $h\to
\gamma\gamma$ looking at the primitive vertices listed in
ref.~\cite{Dedes:2017zog}.  There are 8 classes of such operators
$X^3, \varphi^6, \varphi^4 D^2, \psi^2\varphi^3, X^2\varphi^2$,
$\psi^2 X\varphi, \psi^2 \varphi^2 D, \psi^4$ where $X$ represents a
gauge field strength tensor, $\varphi$ the Higgs doublet, $D$ a
covariant derivative and $\psi$ a generic fermion field.
Not counting flavour multiplicities and hermitian conjugation, in
general, there are 16+2 CP-conserving
operators.\footnote{Incorporating the CP-violating operators will not
  create any problem in the procedure of renormalisation or elsewhere
  in our analysis.  However, these operators are usually strongly
  suppressed by CP-violating type of observables such as particles'
  Electric Dipole Moments (EDMs) and this is the only motivation for
  not considering them in this work. }
Actually, \emph{not all} operators in Table~\ref{tab:no4ferm}
contribute in the final result for the $h\to \gamma\gamma$ amplitude.
The operator $Q_{\varphi}$ cancels out completely after adding all
contributions.  This leaves 17 CP-conserving operators (or Wilson
coefficients) relevant to the $h\to \gamma\gamma$ amplitude.

Another classification of various $d=6$ operators can be devised
alongside with their strength~\cite{Arzt:1994gp,Einhorn:2013kja}.  The
division is between operators that are potentially tree level
generated (PTG operators) and those that are  loop
generated (LG operators) by the more fundamental theory at high
energies (UV-theory) under the assumption that the latter is
\emph{perturbatively decoupled}.  Under this classification, operators
relevant for $h\to \gamma\gamma$ amplitude are arranged as follows:
\begin{table}[t]
\begin{center}
\renewcommand{\arraystretch}{1.2}
\begin{tabular}{||c|c||} 
\hline\hline 
 PTG & LG \\
\hline
$\varphi^6\;\text{and}\;\varphi^4 D^2$ & $X^3$ \\
$\psi^2 \varphi^3$ & $X^2 \varphi^2$ \\
$\psi^2 \varphi^2 D$ & $\psi^2 X  \varphi$\\
$\psi^4$ & \\
\hline\hline
\end{tabular} 
\end{center}
\caption{\sl PTG and LG classes of operators shown in Table~\ref{tab:no4ferm}.}
\label{tab:2}
\end{table}
LG operators are suppressed by $1/(4\pi)^2$ factors for each loop
and may be thought to be sub-dominant corrections with respect to
PTG operators. Relevant to $h\to \gamma\gamma$, PTG and LG classes of
operators are listed in Table~\ref{tab:2}.
On the other hand, a perturbative
decoupling of the UV-theory may not necessarily be the case that
Nature has chosen.  In this work, although we do not \emph{assume} any
distinction amongst the $d=6$ operators involved in $h\to
\gamma\gamma$ amplitude, we shall be referring to Table~\ref{tab:2} as
our analysis progresses.

\section{Renormalisation}
\label{sec:renormalization}

\subsection{Parameter initialisation in SMEFT}

There is a set of very well measured quantities, to which we rely
upon, in relating our calculation for $\mathcal{R}_{h\to \gamma
  \gamma}$ to the LHC data.  This set of experimental values
is~\cite{Patrignani:2016xqp}
\begin{align}
\label{expvalues}
G_F &= 1.1663787(6) \times 10^{-5}\;\text{GeV}^{-2} \,, \nonumber\\
\alpha_{\mathrm{EM}} &= 1/137.035999139(31) \quad\text{at}\; Q^2=0\,,  \nonumber\\
M_W &= 80.385(15)\;\text{GeV} \,, \nonumber  \\
M_Z &= 91.1876(21)\;\text{GeV}\,, \nonumber  \\
M_h &= 125.09 \pm 0.24\;\text{GeV}\,, \nonumber \\
m_t &= 173.1 \pm 0.6\;\text{GeV} \,.
\end{align}
We identify these input values with the ones obtained in SMEFT
consistent with the given accuracy of up to $1/\Lambda^2$ expansion
terms.
Consequently, following ref.~\cite{Dedes:2017zog} for the gauge and
Higgs boson masses at tree level, it is enough to set $M_W$, $M_Z$ and
$M_h$, respectively, equal to
\begin{align}
M_{W}&= \frac{1}{2} \bar g v \,,\nonumber\\
M_{Z}&= \frac{1}{2} \sqrt{\bar g^2 + \bar g^{\prime 2}} v \left(1 +
\frac{ \bar g \bar g' C^{\varphi WB} v^2}{\bar g^2 + \bar
g^{\prime 2}} + {1\over 4}C^{\varphi D}v^2\right)\,,\nonumber\\
M_{h}^{2}&= \lambda v^2 - \left( 3 C^{\varphi} - 2 \lambda
C^{\varphi\square} + \frac{\lambda}{2} C^{\varphi D} \right) v^4 \,,
\end{align}
where $\lambda$ is the Higgs quartic coupling, $\bar g', \bar g$ are,
respectively, the $U(1)_Y$ and $SU(2)_L$ gauge couplings (redefined to
obtain canonical form of the gauge kinetic terms, see
ref.~\cite{Dedes:2017zog}) and the $C$-coefficients correspond to
operators defined in Table~\ref{tab:no4ferm}.  Moreover, the fine-structure constant is identified through the Thomson limit ($Q^2=0$)
as $\alpha_{\mathrm{EM}} = \bar{e}^2/4\pi$ where $\bar{e}$ is given at
tree level by
\begin{align}
\label{eq:ebar}
\bar{e}=\frac{\bar{g} \bar{g}^{\prime}}{\sqrt{\bar{g}^2 +
    \bar{g}^{\prime 2}}} \left(1 - \frac{ \bar g \bar g' }{\bar g^2 +
  \bar g^{\prime 2}} \, C^{\varphi WB}\, v^2\right)\,.
\end{align}
Similarly, the experimental values for lepton and quark masses, taken
as pole masses from ref.~\cite{Patrignani:2016xqp}, are equal to
eqs.~(3.27) and~(3.29) of ref.~\cite{Dedes:2017zog}.

The Fermi coupling constant $G_F$, is identified through the muon
decay process.  In addition to the $W$-boson exchange which is
modified in SMEFT by the PMNS matrix that is (now) \emph{a
  non-unitary} matrix containing the coefficient $C_{\varphi l}^{(3)}$,
$G_F$ is also affected by dipole operators e.g., $Q_{eW}$ or by new
diagrams with $Z$- or Higgs-boson exchange.  However, the expression
for $G_F$ is simplified by making the approximation of zero neutrino
masses and also by \emph {assuming} that
\begin{equation}
C_1\, {v^2} \gg C_2 \, {v\, m_l} \,,
\label{Gfapprox}
\end{equation}
for any generic $C_1$ and $C_2$ coefficients entering the muon-decay
amplitude and $m_l$ being a charged lepton mass.  Only then we
identify the Fermi coupling constant of \eq{expvalues}, within
\emph{tree level} in SMEFT, as
\begin{equation}
\frac{G_{F}}{\sqrt{2}}=\frac{\bar{G}_F}{\sqrt{2}} \left [ 1 + v^2 (
  C_{11}^{\varphi l (3)} + C_{22}^{\varphi l (3)} ) - v^2
  C_{1221}^{ll} \right ] \,, \qquad \text{with} \quad
\frac{\bar{G}_F}{\sqrt{2}} \equiv \frac{\bar{g}^2}{8 M_W^2} =
\frac{1}{2 v^2} \,.
\label{eq:GF}
\end{equation}
All Wilson coefficients entering in eq.~\eqref{eq:GF} are real since
they are diagonal elements of Hermitian matrices.  In fact, and as a
side test of the approximations assumed in eq.~\eqref{Gfapprox}, we
have checked that, at tree level in SMEFT, the full $S$-matrix element
for the process $\mu^- \to e^- \bar{\nu}_e \nu_\mu$ is gauge invariant
independently of lepton-number conservation.  The formula
\eqref{eq:GF} agrees with the corresponding one from
refs.~\cite{Brivio:2017vri, Dawson:2018pyl}.

\subsection{Renormalisation framework}

We ultimately want to bring the expression for the amplitude
$\mathcal{A}(h\to \gamma\gamma)$, into a form that contains only
renormalised parameters that are most closely related to observable
quantities, the relevant ones given in \eq{expvalues}.  At tree level
in SMEFT, the $h\gamma\gamma$-vertex appears only in association with
the unrenormalised (bare) Wilson coefficients, $C^{\varphi B}_0,
C^{\varphi W}_0$ and $C^{\varphi WB}_0$ and these are multiplied by
the bare vev parameter\footnote{In fact this is $\bar{v}_0$ but to
  order $1/\Lambda^2$ it is replaced with the ``unbarred" parameter,
  $v_0$.} $v_0$ (in what follows bare parameters are always denoted
with a subscript zero).  In order to set the stage, let us for example
consider from Table~\ref{tab:no4ferm} the $d=6$, CP-invariant operator
of the form $X^2\varphi^2$,
\begin{equation}
C^{\varphi B}_0\, \varphi^\dagger \varphi \, B_{\mu\nu} \,
B^{\mu\nu}\,,
\label{eq:ffBB}
\end{equation}
where $\varphi$ is the scalar Higgs doublet and $B_{\mu\nu}$ is the
$U(1)_Y$-hypercharge gauge field strength tensor.  All fields and
coupling constants are unrenormalised quantities in this expression.
In what follows, and in order to keep the expressions as simple as
possible, we keep working with unrenormalised fields {\it i.e.,} no
usual field redefinition is performed.  This is justified, because we
are interested in calculating only an on-shell $S$-matrix amplitude rather than
a Green function.\footnote{This is more important than, as it sounds,
  just a calculational scheme.  Certain operators vanish when using
  equations of motion.  Green functions are affected by these
  operators whereas their $S$-matrix elements
  vanish~\cite{Politzer:1980me,Arzt:1993gz,Georgi:1991ch}.}

After Spontaneous Symmetry Breaking (SSB) in SMEFT (see
ref.~\cite{Dedes:2017zog} for details), the expression in
eq.~\eqref{eq:ffBB} contains the following term describing the
interaction of the Higgs field and two ``photons'',
\begin{equation}
C_0^{\varphi B} \, v_0 \, h \, B_{\mu\nu} \, B^{\mu\nu}\,,
\label{eq:ffBBvac}
\end{equation}
where $h$ is the Higgs field.
We split these bare quantities into renormalised parameters
$v,C^{\varphi B}$ and counterterms, $ \delta v, \delta C^{\varphi B}$
respectively, as
\begin{equation}
v_0=v - \delta v \,, \qquad C^{\varphi B}_0= C^{\varphi B}  -  \delta
C^{\varphi B}\,.
\end{equation}
We follow the steps of a simple on-shell renormalisation scheme, first
described in SM by Sirlin~\cite{Sirlin:1980nh}, and introduce new
unrenormalised fields $A_\mu$ and $Z_\mu$ through the linear
combinations
\begin{align}
	B_\mu &= c A_\mu - s Z_\mu \,, \\
    W_\mu^3 &=s A_\mu+ c Z_\mu \,,
\end{align} 
with $c\equiv \cos\theta_W$ and $s\equiv \sin\theta_W$ defined as a
ratio of the physical masses of $W$ and $Z$ bosons, like
\begin{equation}
  c^2\equiv \cos^2\theta_W=\frac{M_W^2}{M_Z^2} \,. \label{eq:c}
\end{equation}
Therefore, the Lagrangian term for the considered operator,
$Q_{\varphi B}$, describing (part of) the $h\gamma\gamma$ interaction,
reads,
\begin{equation}
c^2 \, v \, C^{\varphi B}\, \left [1 - \frac{\delta C^{\varphi
      B}}{C^{\varphi B}} - \frac{\delta v}{v} \right ] \, h \,
F_{\mu\nu} F^{\mu\nu} \,.
\end{equation}
Note that the vev counterterm arises from pure SM contributions
because it multiplies $C^{\varphi B}$, while $\delta C^{\varphi B}$
cancels infinities that arise only from pure SMEFT diagrams \ie in
general, diagrams proportional to other $C$-coefficients, not
necessarily only $C^{\varphi B}$.

\begin{figure}
\centerline{\includegraphics[scale=0.5]{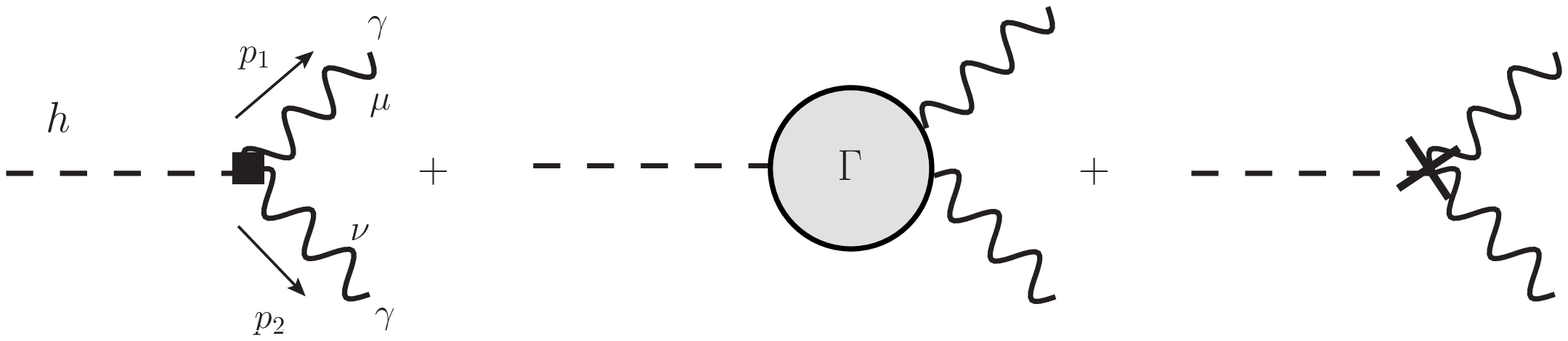}}
\caption{\sl The sum of three types of diagrams: (left) the SMEFT
  ``tree" contribution with momenta and space-time indices indicated,
  (center) the 1PI vertex corrections $\Gamma$ from all operators, SM
  or not, and (right) the vertex counterterms containing $\delta C$
  and $\delta v$.  These corrections should be self-explained in
  eq.~\eqref{m1}.}
\label{fig1-1PI}
\end{figure}

Besides operator $Q_{\varphi B}$, counterterms for operators
$Q_{\varphi W}$ and $Q_{\varphi WB}$ need to be added, too.  Because
all these three operators are proportional to the Higgs bilinear
combination, $\varphi^\dagger \varphi$, they all contain the vev
  counterterm as a universal contribution to $h\to \gamma\gamma$
amplitude.  The contributions discussed so far are depicted and
explained in Fig.~\ref{fig1-1PI}.  By making use of the Feynman rules
of ref.~\cite{Dedes:2017zog}, their sum is written in momentum space,
as
\begin{align}
4 i\, [\, p_1^\nu\, p_2^\mu - (p_1\cdot p_2)\, g^{\mu\nu} \, ] \,
\Biggl \{& c^2\, v\, C^{\varphi B} \left [ 1 + \Gamma^{\varphi B} -
  \frac{\delta C^{\varphi B}}{C^{\varphi B}} - \frac{\delta v}{v}
  \right ] \nonumber \\
& + s^2 \, v \, C^{\varphi W} \left [ 1 + \Gamma^{\varphi W} -
  \frac{\delta C^{\varphi W}}{C^{\varphi W}} - \frac{\delta v}{v}
  \right ] \nonumber \\
& - s c \, v \, C^{\varphi WB} \left [ 1 + \Gamma^{\varphi WB} -
  \frac{\delta C^{\varphi WB}}{C^{\varphi WB}} - \frac{\delta v}{v}
  \right ] \nonumber \\
& + \frac{1}{M_W} \overline{\Gamma}^{\mathrm{SM}} + \sum_{X \ne
  \varphi B, \varphi W, \varphi WB} v \, C^X \, \Gamma^X \Biggr \}\,.
\label{m1}
\end{align}
One-loop, 1PI vertex contributions proportional to $C^{\varphi B},
C^{\varphi W}$ and $C^{\varphi WB}$ are denoted (up to pre-factors)
with $\Gamma^{\varphi B}, \Gamma^{\varphi W}$ and $\Gamma^{\varphi
  WB}$ in the first three lines of the above equation.  The SM
contribution, $\overline{\Gamma}^{\mathrm{SM}}$, is just the SM-famous
result of ref.~\cite{Ellis:1975ap} but with the SM parameters replaced
by the SMEFT ones (that is why ``barred'' $\Gamma$), taken from
refs.~\cite{Dedes:2017zog, Alonso:2013hga}.  Furthermore, there are
additional one-loop corrections, $\Gamma^X$, proportional to Wilson
coefficients $C^X$, like for instance $C^W$, which are collected in
the last line, last term of \eq{m1}.

\begin{figure}
\centerline{\includegraphics[scale=0.55]{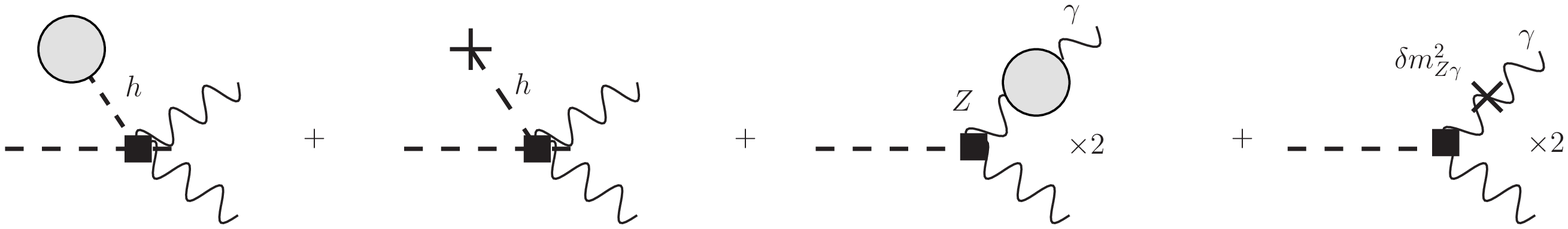} }
\caption{\sl Tadpole and $Z\gamma$ self-energy contributions with
  their associated counterterms.  Crosses denote SM counterterms and
  the black boxes indicate pure $d=6$ operator insertions.}
\label{fig2:tadsZg}
\end{figure}

There are additional diagrams participating in the $h\to \gamma\gamma$
amputated amplitude.  These are shown in Fig.~\ref{fig2:tadsZg}.  The
first two classes of diagrams are the Higgs tadpole and its
counterterm contributions.  These two diagrams do not enter in our
renormalised amplitude because, following the renormalisation scheme
of ref.~\cite{Sirlin:1985ux}, the Higgs tadpole counterterm is
adjusted to cancel the 1PI Higgs tadpole diagrams.  This guarantees
that the vev is unchanged to one-loop order.
The last two diagrams in Fig.~\ref{fig2:tadsZg} represent the
$Z\gamma$-self energy at $p^2=0$, $A_{Z\gamma}(0)$, plus its
counterterm, $\delta m_{Z\gamma}^2$.  The expression for the
counterterm $\delta m^2_{Z\gamma}$ (given below) is gauge invariant
independently of the renormalisation condition for the Higgs tadpole.
This is practically very useful for proving the gauge invariance of
the $h\to \gamma\gamma$ amplitude.

\begin{figure}
\centerline{
  \includegraphics[scale=0.55]{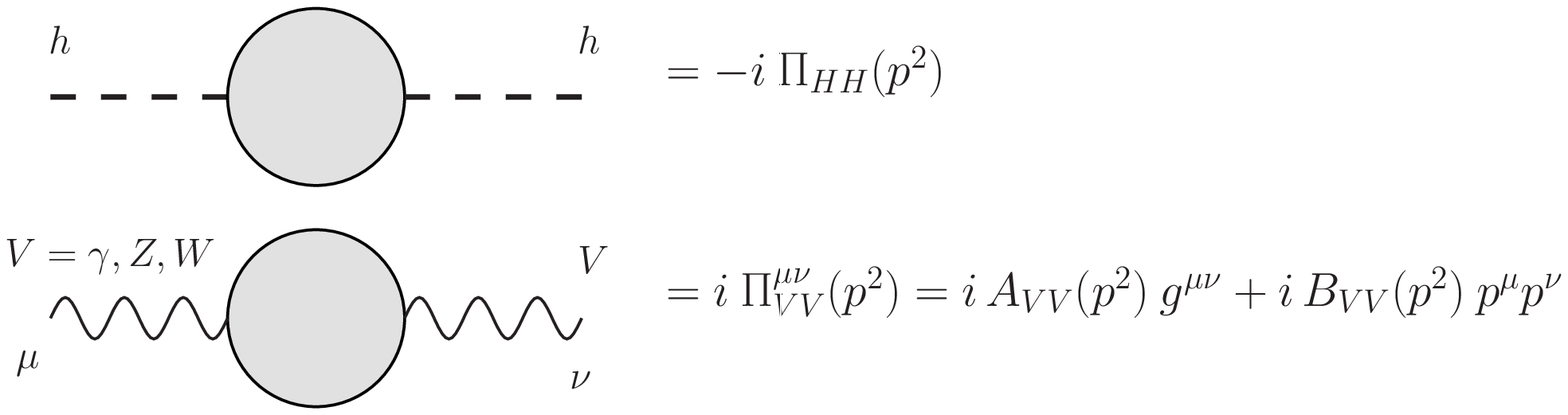} }
 \caption{\sl Definitions for Higgs and vector boson ($V=\gamma,
   Z,W$), 1PI self-energies.
\label{fig:se}}
\end{figure}

Finally, as usual, by multiplying the amputated graph with the
LSZ-factors~\cite{Lehmann:1954rq} (see for instance section 7.2 of
textbook~\cite{Peskin:1995ev}) for the external Higgs and photon
fields,
\begin{equation}
\sqrt{Z_{hh}} \, Z_{\gamma\gamma} = 1 + \frac{1}{2} \Pi_{HH}^\prime(M_h^2)
- \Pi_{\gamma\gamma}(0) \,,
\end{equation}
we arrive at the following $S$-matrix amplitude:
\begin{align}
&i \mathcal{A}^{\mu\nu}(h\to \gamma\gamma) = \bra{\gamma
    (\epsilon^\mu, p_1), \,\gamma (\epsilon^\nu, p_2)\,}\, S \,
  \ket{\, h(q)\, } = 4 i \, \left[ \, p_1^\nu\, p_2^\mu - (p_1\cdot
    p_2)\, g^{\mu\nu} \, \right] \times\nonumber \\[2mm]
& \Biggl \{ c^2\, v\, C^{\varphi B} \left [ 1 + \Gamma^{\varphi B} -
    \frac{\delta C^{\varphi B}}{C^{\varphi B}} - \frac{\delta v}{v} +
    \frac{1}{2}\Pi_{HH}^\prime (M_h^2) - \Pi_{\gamma\gamma}(0) + 2
    \tan\theta_W \frac{A_{Z\gamma}(0) + \delta
      m^2_{Z\gamma}}{M_Z^2} \right ] \nonumber \\[2mm]
& + s^2 \, v \, C^{\varphi W} \left [ 1 + \Gamma^{\varphi W} -
    \frac{\delta C^{\varphi W}}{C^{\varphi W}} - \frac{\delta v}{v} +
    \frac{1}{2}\Pi_{HH}^\prime (M_h^2) - \Pi_{\gamma\gamma}(0) -
    \frac{2}{\tan\theta_W} \frac{A_{Z\gamma}(0) + \delta
      m^2_{Z\gamma}}{M_Z^2} \right ] \nonumber \\[2mm]
& - s c \, v \, C^{\varphi WB} \left [ 1 + \Gamma^{\varphi WB} -
    \frac{\delta C^{\varphi WB}}{C^{\varphi WB}} - \frac{\delta v}{v}
    + \frac{1}{2}\Pi_{HH}^\prime (M_h^2) - \Pi_{\gamma\gamma}(0) -
    \frac{2}{\tan 2\theta_W} \frac{A_{Z\gamma}(0) + \delta
      m^2_{Z\gamma}}{M_Z^2} \right ] \nonumber \\[2mm]
& + \frac{1}{M_W} \overline{\Gamma}^{\mathrm{SM}} +
  \sum_{X \ne \varphi B, \varphi W, \varphi WB} \, v \, C^X \,
  \Gamma^X \Biggr \}\,.
\label{amp}
\end{align}
Eq.~(\ref{amp}) is our master formula for the renormalised amplitude
$\mathcal{A}^{\mu\nu}(h\to \gamma\gamma)$.  The definitions for the
various self-energies\footnote{We follow closely the notation of
  ref.~\cite{Sirlin:1980nh}.} are stated in Fig.~\ref{fig:se} and
\begin{equation}
\Pi^\prime_{HH}(M_h^2) \equiv \frac{\partial \Pi_{HH}(p^2)}{\partial
  p^2} \biggr |_{p^2=M_h^2} \,, \qquad A_{\gamma\gamma}(p^2) = - p^2
\, \Pi_{\gamma\gamma}(p^2) + \mathcal{O}(\alpha_{\mathrm{EM}}^2) \,,
\end{equation}
where $\Pi_{\gamma\gamma}(p^2)$ is regular at $p^2=0$.  All
self-energies in \eq{amp} should arise purely from SM diagrams because
we are including terms up to $1/\Lambda^2$ in SMEFT.  As noted
earlier, the SM counterterm, $\delta m^2_{Z\gamma}$, is {\em gauge
  invariant} and is given by~\cite{Sirlin:1980nh}:
\begin{equation}
\frac{\delta m^2_{Z\gamma}}{M_{Z}^2} = \frac{1}{2 \tan\theta_W}
\operatorname{Re}\left[ \frac{A_{ZZ}(M_Z^2)}{M_Z^2} -
  \frac{A_{WW}(M_W^2)}{M_W^2} \right]\,.
 \label{dmza}
\end{equation}
The quantity $\delta v/v$ is not gauge invariant.  Following standard
on-shell renormalisation conditions of refs.~\cite{Sirlin:1980nh,
  Sirlin:1985ux}, we write
\begin{eqnarray}
\frac{\delta v}{v} = \operatorname{Re}\left[ \frac{A_{WW}(M_W^2)}{2
    M_W^2}\right] - \frac{\delta g}{g}\,,
\label{vevc}
\end{eqnarray} 
where the counterterm $\delta g$ of the $SU(2)_L$ gauge coupling is
{\em gauge invariant} and reads as
\begin{equation}
\frac{\delta g}{g} = \frac{\delta e}{e} - \frac{1}{\tan\theta_W} \,
\frac{\delta m^2_{Z\gamma}}{M_{Z}^2} \,.
\end{equation}
Here $\delta e$ is the electromagnetic charge renormalisation
counterterm which is also {\em gauge invariant}.  This is given by
eq.~(26) of ref.~\cite{Sirlin:1980nh}
\begin{equation}
\frac{\delta e}{e} = - \frac{1}{2}
\Pi^{\mathrm{lept}}_{\gamma\gamma}(0) - \frac{1}{2}
\Pi^{\mathrm{had}}_{\gamma\gamma}(0) + \frac{7 e^2}{32 \pi^2} \left[
  \left ( \frac{2}{\epsilon} - \gamma + \log 4 \pi \right ) - \log
  \frac{M_W^2}{\mu^2} + \frac{2}{21} \right]\,,
  \label{eq:dee}
\end{equation}  
where $\mu$ is the renormalisation scale parameter and $\epsilon\equiv
4 - d$.  Leptonic and hadronic contributions,
$\Pi^{\mathrm{lept}}_{\gamma\gamma}(0)$ and
$\Pi^{\mathrm{had}}_{\gamma\gamma}(0)$, to the photon vacuum
polarisation are gauge invariant and the infinite part in the squared
brackets should be gauge invariant too.  The hadronic contribution
from light quarks, $\Pi^{\mathrm{had}}_{\gamma\gamma}(0)$, is in
principle non-calculable due to strong interaction at zero momenta.  A
dispersive or other non-perturbative methods should be in order.
There is no such problem of course with
$\Pi^{\mathrm{lept}}_{\gamma\gamma}(0)$.

SM vector boson self-energy contributions can be found in
ref.~\cite{Marciano:1980pb}.  The Higgs self-energy contribution can
be found in refs.~\cite{Sirlin:1985ux,Hartmann:2015oia}.  These
results have been obtained in the particular case of the 't
Hooft-Feynman gauge where $\xi=1$.  Thanks to the set of SMEFT Feynman
Rules in general $R_\xi$-gauges~\cite{Dedes:2017zog}, we present in
Appendix~\ref{app:pv} all contributions needed in \eq{amp} with the
explicit $\xi$-dependence.  This is necessary for checking the gauge
invariance of the amplitude.  Finally, the counterterms $\delta
C^{\varphi B}, \delta C^{\varphi W}$ and $\delta C^{\varphi WB}$ can
be read from refs.~\cite{Grojean:2013kd, Jenkins:2013zja,
  Jenkins:2013wua, Alonso:2013hga, Hartmann:2015oia, Hartmann:2015aia}
where they have been calculated again in 't Hooft-Feynman ($\xi=1$)
gauge.  However, in $\overline{\mathrm{MS}}$ renormalisation scheme
and at one-loop, cancellation of infinities should be independent on
the gauge choice as we confirm below.
 
\subsection{$\xi$-independence}

Knowing the gauge invariant and non-invariant parts of various
contributions, as described above, is particularly useful for proving
the $\xi$-independence of the amplitude.  We first prove gauge
invariance by means of $\xi$-independence for the infinite parts
proportional to $\xi_W$ or $\xi_Z$.
We find that the combination of $\delta v/v$ and $\Pi_{HH}^\prime(M_h^2)$
in eq.~\eqref{amp} is $\xi$-independent.
For the $C^{\varphi B}$ contribution in eq.~\eqref{amp}, the
$\xi_W$-dependent terms inside $\Pi_{\gamma\gamma}(0)$ and
$A_{Z\gamma}(0)$ cancel among each other, as they should since the infinite part
of $\Gamma^{\varphi B}$ is $\xi$-independent by itself.
For contributions proportional to $C^{\varphi W}$ ($C^{\varphi WB}$),
the $\xi_W$ cancellations take place throughout the
self-energy contributions and 
$\Gamma^{\varphi W}$ ($\Gamma^{\varphi W B}$).
Furthermore, diagrams proportional to $C^X$ with $X \ne \varphi B, \varphi W,
\varphi WB$, contributing to the last term of \eq{amp}, are gauge
invariant on their own.  Of course $\overline{\Gamma}^{\mathrm{SM}}$
is finite and gauge invariant as it is known from a direct calculation
in $R_\xi$-gauges with dimensional
regularisation~\cite{Marciano:2011gm}.\footnote{For a strict
  four-dimensional calculation in \emph{unitary} gauge, see
  ref.~\cite{Dedes:2012hf}.}

We then prove analytically the cancellation of all $\xi$-dependent
finite parts.  This was done by first performing a maximal reduction
on the related Passarino-Veltman functions~\cite{Passarino:1978jh} and
then analytically checking for $\xi$-dependence among the parametric
integrals.  This is a highly non-trivial check of the validity of our
calculation because the gauge parameter $\xi$ appears everywhere in
\emph{both} the SM and SMEFT contributions which are directly related
to the $h\to \gamma\gamma$ amplitude.
Moreover, this should be also considered as a direct proof for the
validity of the expressions for vertices given in
ref.~\cite{Dedes:2017zog} in general $R_\xi$-gauges.
Most importantly, the $\xi$-cancellation shows that the amplitude
$\mathcal{A}^{\mu\nu}(h\to \gamma\gamma)$ given in \eq{amp} is {\em
  gauge invariant} as it should be.  Needless to say, this is a very
encouraging indication towards the correctness of our final result.

As an additional non-trivial check of our calculation, we have also
proved gauge invariance for our amplitude before adopting any
renormalisation scheme.  We confirm that the regularised but yet
  unrenormalised $S$-matrix amplitude for $h\to \gamma\gamma$,
written in terms of bare parameters, is {\em gauge invariant}.

\subsection{$\overline{\mathrm{MS}}$ scheme for Wilson coefficients}

All renormalised coefficients, say $C$, and the counterterms, $\delta
C$, in eq.~\eqref{amp}, can be readily written in terms of the
$\overline{\mathrm{MS}}$-scheme running $C$-coefficients as
\begin{equation}
C - \delta C = \bar{C}(\mu) - \delta \bar{C} \,,
\end{equation}
where $\mu$ is the renormalisation (or subtraction) scale that lays
somewhere between the EW scale and the scale $\Lambda$, while $\delta
\bar{C}$ is a counterterm that subtracts only terms proportional to
\begin{equation}
E \equiv \frac{2}{\epsilon} - \gamma + \log 4 \pi \,, \quad
\mathrm{with} \quad \epsilon \equiv 4 - d \,,
\label{eq:e}
\end{equation}
in the loop corrections for the Wilson $C$-coefficients.  In
$\overline{\mathrm{MS}}$ scheme and at one-loop, these counterterms
are independent of the choice of the gauge fixing and can be read
directly from refs.~\cite{Jenkins:2013zja, Alonso:2013hga,
  Jenkins:2013wua} to be
\begin{align}
\delta \bar{C}^{\varphi B}=\frac{E}{16\pi^2} \Biggl \{ & \biggl ( -
3 \lambda - Y + \frac{9}{4} \bar{g}^2 - \frac{85}{12} \bar{g}^{\prime 2} \biggr )
C^{\varphi B} - \frac{3}{2} \bar{g} \bar{g}^\prime C^{\varphi WB} \nonumber\\&
- \left [ \frac{3}{2} \bar{g}^\prime \operatorname{Tr}(C^{\prime eB}
  \Gamma_e^\dagger) - \frac{5}{6} \bar{g}^\prime N_c
  \operatorname{Tr}(C^{\prime uB} \Gamma_u^\dagger) + \frac{1}{6}
  \bar{g}^\prime N_c \operatorname{Tr}(C^{\prime dB} \Gamma_d^\dagger)
  + \mathrm{H.c.} \right ] \Biggr \}\,,
\label{eq:dCphiB} 
\end{align}
\begin{align}
\delta \bar{C}^{\varphi W}=\frac{E}{16\pi^2} \Biggl \{ & \Biggl ( - 3
\lambda - Y + \frac{53}{12} \bar{g}^2 + \frac{3}{4} \bar{g}^{\prime 2}
\Biggr ) C^{\varphi W} - \frac{1}{2} \bar{g} \bar{g}^\prime C^{\varphi
  W B} + \frac{15}{2} \bar{g}^3 C^W \nonumber \\&
+\left [ \frac{1}{2}\bar{g}\, \operatorname{Tr}(C^{\prime eW}
  \Gamma_e^\dagger) + \frac{1}{2} \bar{g}\, N_c
  \operatorname{Tr}(C^{\prime uW} \Gamma_u^\dagger) + \frac{1}{2}
  \bar{g}\, N_c \operatorname{Tr}(C^{\prime dW} \Gamma_d^\dagger) +
  \mathrm{H.c.}\right ] \Biggr \}\,, \label{eq:dCphiW}
\end{align}
\begin{align}
{\delta \bar{C}^{\varphi WB}}=\frac{E}{16\pi^2} \Biggl \{& \Biggl ( -
\lambda - Y - \frac{2}{3} \bar{g}^2 - \frac{19}{6} \bar{g}^{\prime 2}
\Biggr ) C^{\varphi W B} - \bar{g} \bar{g}^\prime ( C^{\varphi B} +
C^{\varphi W} ) - \frac{3}{2} \bar{g}^\prime \bar{g}^2 C^W \nonumber
\\&
+ \left [ \frac{1}{2}\bar{g}\, \operatorname{Tr}(C^{\prime eB}
  \Gamma_e^\dagger) -\frac{1}{2} \bar{g}\, N_c
  \operatorname{Tr}(C^{\prime uB} \Gamma_u^\dagger) + \frac{1}{2}
  \bar{g}\, N_c \operatorname{Tr}(C^{\prime dB} \Gamma_d^\dagger)
  \right.\nonumber \\& \left.
- \frac{3}{2} \bar{g}^\prime \operatorname{Tr}(C^{\prime eW}
\Gamma_e^\dagger) - \frac{5}{6} \bar{g}^\prime N_c
\operatorname{Tr}(C^{\prime uW} \Gamma_u^\dagger) - \frac{1}{6}
\bar{g}^\prime N_c \operatorname{Tr}(C^{\prime dW} \Gamma_d^\dagger) +
\mathrm{H.c.}  \right ] \Biggr \}\,,\label{eq:dCphiWB}
\end{align}
where $\Gamma_{u,d,e}$ is our
notation~\cite{Grzadkowski:2010es,Dedes:2017zog} for the usual Yukawa
couplings in SM, and using Table~4 from \Ref{Dedes:2017zog}, the
coefficients $C^{\prime\, f}$ are rotated to the fermion mass basis
(denoted now as unprimed ones), and
\begin{equation}
Y \equiv \frac{2}{v^2} \sum_{i=1}^3 (m_{e_i}^2 + N_c m_{d_i}^2 + N_c
m_{u_i}^2 ) \,, \qquad 
\operatorname{Tr}(C^{\prime eB} \Gamma_e^\dagger) =
\frac{\sqrt{2}}{v} C^{eB}_{ii} m_{e_i}\,, \quad \text{etc.}
\end{equation}
$N_c=3$ is the number of colours and $m_{f_i}$ a mass of the SM
fermion belonging to the $i$-th generation.  All $C$-coefficients have
been taken real.
We have checked explicitly and analytically that the counterterms of
\eqss{eq:dCphiB}{eq:dCphiW}{eq:dCphiWB} render the amplitude for $h\to
\gamma\gamma$ of \eq{amp} \emph{finite}, at one-loop and up to
$1/\Lambda^2$ in EFT expansion.

\subsection{The amplitude}

The remaining part of $\mathcal{A}^{\mu\nu}(h\to \gamma\gamma)$ in
eq.~\eqref{amp} is, at one-loop and up to $1/\Lambda^2$ terms,
renormalisation scale invariant: the renormalisation group running of
$\bar{C}(\mu)$ coefficients cancels the explicit $\mu$-dependence
within various contributions in the RHS of \eq{amp}.  Therefore, the
amplitude, to be squared in finding the $h\to \gamma\gamma$ decay
width, is
\begin{equation}
i \mathcal{A}^{\mu\nu}(h\to \gamma\gamma) = \bra{\gamma
  (\epsilon^\mu, p_1), \,\gamma (\epsilon^\nu, p_2)\,}\, S \,
\ket{\, h(q)\, } = 4 i \, \left[\, p_1^\nu\, p_2^\mu - (p_1\cdot
  p_2)\, g^{\mu\nu} \, \right] \, \mathcal{A}_{h\to \gamma \gamma} \,,
\label{physamp2}
\end{equation}
where 
\begin{align}
\mathcal{A}_{h\to \gamma \gamma} &= \Biggl \{ c^2\, v\,
\bar{C}^{\varphi B}(\mu) \left [ 1 + \Gamma^{\varphi B} - \frac{\delta
    v}{v} + \frac{1}{2}\Pi_{HH}^\prime (M_h^2) - \Pi_{\gamma\gamma}(0)
  + 2 \tan\theta_W \frac{A_{Z\gamma}(0) + \delta
    m^2_{Z\gamma}}{M_Z^2} \right ] \nonumber \\[2mm] &\qquad
    + s^2 \, v \, \bar{C}^{\varphi W}(\mu) \left [ 1 + \Gamma^{\varphi W} -
\frac{\delta v}{v} + \frac{1}{2}\Pi_{HH}^\prime (M_h^2) -
\Pi_{\gamma\gamma}(0) - \frac{2}{\tan\theta_W} 
\frac{A_{Z\gamma}(0) + \delta m^2_{Z\gamma}}{M_Z^2} \right ]
\nonumber \\[2mm] &\qquad
 - s c \, v \, \bar{C}^{\varphi WB}(\mu) \left [ 1 + \Gamma^{\varphi
    WB} - \frac{\delta v}{v} + \frac{1}{2}\Pi_{HH}^\prime (M_h^2) -
  \Pi_{\gamma\gamma}(0) - \frac{2}{\tan 2\theta_W} 
  \frac{A_{Z\gamma}(0) + \delta m^2_{Z\gamma}}{M_Z^2} \right
] \nonumber \\[2mm]&\qquad
 + \frac{1}{M_W} \overline{\Gamma}^{\mathrm{SM}} +
\ \sum_{X \ne \varphi B, \varphi W, \varphi WB} \, v \, C^X(\mu) \,
\Gamma^X \, \Biggr \}_{\mathrm{finite}} \,.
\label{physamp}
\end{align}
The subscript ``finite" in the final parenthesis means that infinities
proportional to $E$ have been subtracted from all contributions in
eq.~\eqref{physamp} such as $\Gamma$, $\Pi^\prime_{HH}$, $\Pi_{VV}$,
$A_{VV}$, \emph{etc}.
The $\mathcal{A}_{h\to \gamma \gamma}$ in eq.~\eqref{physamp} is
finite, gauge and renormalisation scale invariant\footnote{In the
  sense that ${d \over d\mu} \mathcal{A}_{h\to\gamma\gamma}(\mu)=0$.}
as a physical amplitude must be.  In eq.~\eqref{physamp},
$\Gamma^{\varphi B}$, $\Gamma^{\varphi W}$ and $\Gamma^{\varphi WB}$
are given in Appendix~\ref{app:pv} in eqs.~\eqref{CB}, \eqref{CW} and
\eqref{CWB}.  The quantities ${\delta v}/{v}$ and ${\delta
  m^2_{Z\gamma}}/{M_Z^2}$ are presented in eqs.~\eqref{vevc} and
\eqref{dmza}, respectively.  All vector boson self-energies in general
$R_\xi$-gauges as well as the quantity $\Pi_{HH}^\prime (M_h^2)$ are
also given in Appendix~\ref{app:pv}.

Although all $\bar{C}(\mu)$ coefficients in eq.~\eqref{physamp} are
$\overline{\mathrm{MS}}$ parameters, the weak mixing angle $\theta_W$
and the vev $v$ that appear explicitly to multiply Wilson coefficients
are defined in terms of physical quantities through
eqs.~\eqref{eq:c} and \eqref{eq:GF} [see also eq.~\eqref{eq:GFcor}
  below].  This is a virtue of our hybrid renormalisation scheme: SM
on-shell parameters appear together with $\overline{\mathrm{MS}}$
SMEFT parameters (Wilson coefficients) in the renormalised amplitude.
This scheme can easily be applied to every process at one-loop in
SMEFT.

From now on, all Wilson coefficients should be considered as running
$\overline{\mathrm{MS}}$ quantities, $C \equiv \bar{C}(\mu)$.  We
remove the ``bar'' over the $\overline{\mathrm{MS}}$-coefficients
letting the argument to denote, or to implicitly imply, the
difference.

\section{Anatomy of the effective amplitude}
\label{sec:amplitude}

In this section we present explicit expressions for the SM
contribution, and, contributions proportional to all Wilson
coefficients entering the $h\to \gamma\gamma$ amplitude in
eq.~\eqref{physamp}, and  in Table~\ref{tab:no4ferm}.  These
coefficients are taken to be real.  For clarity, we reinstate
explicitly $1/\Lambda^2$ factors in the expressions appeared in this
and subsequent sections, so they are no longer incorporated into the
definition of $C$'s. Our EFT expansion stops at the order
$1/\Lambda^2$ and is one-loop at the $\hbar$-expansion.
In our conventions, we denote electromagnetic fermion charges and the
third component of particle weak isospin as
\begin{equation}
Q_{f}=
\begin{cases}
0,&\text{for} \quad f=\nu_{e},\nu_{\mu},\nu_{\tau}\\
 - 1,&\text{for} \quad f=e,\mu,\tau\\
2/3,&\text{for} \quad f=u,c,t\\
 - 1/3,\;&\text{for} \quad f=d,s,b\\
\end{cases}\quad \text{and} \quad
T_{f}^{3}=
\begin{cases}
1/2,&\text{for} \quad f=\nu_{e},\nu_{\mu},\nu_{\tau},u,c,t\\
 - 1/2,\;&\text{for} \quad f=e,\mu,\tau,d,s,b
\end{cases}\,.
\label{eq:charges}
\end{equation}
The colour factors are $N_{c,e}=1$ and $N_{c,u}=N_{c,d}=3$.
It is useful to note, when reading the expressions below, that the
actual dimensionless EFT expansion parameter is $\frac{1}{G_F
  \Lambda^2}$.  To get a quantitative feeling of its numerical
magnitude and to compare with standard loop expansion in the EW gauge
couplings, we simply note that it is $\frac{1}{G_F M_W^2} \sim 4 \pi$,
while for $\Lambda=1~\mathrm{TeV}$ one has $\frac{1}{G_F
  \Lambda^2}\sim \frac{1}{4\pi}$, for $\Lambda=10~\mathrm{TeV}$ one
has $\frac{1}{G_F \Lambda^2}\sim \frac{\alpha_{\mathrm{EM}}}{4\pi}$
and, finally, for $\Lambda=100~\mathrm{TeV}$ one has $\frac{1}{G_F
  \Lambda^2}\sim \frac{\alpha^2_{\mathrm{EM}}}{\pi^2}$.

\subsection{SM and $C^{\varphi WB}$, $C^{\varphi l (3)}$, $C^{l l}$}

The famous ``SM'' contributions from $W$ and fermion triangle loops
are represented by the penultimate term in eq.~\eqref{physamp}.  This
is
\begin{equation}
\frac{\overline{\Gamma}^{\mathrm{SM}}}{M_W} = \frac{1}{64\pi^2}
\frac{\bar{g}^2 \bar{g}^{\prime 2}}{(\bar{g}^2 + \bar{g}^{\prime 
    2})} \frac{\bar{g}}{M_W} I_{\gamma\gamma}\,,
\label{eq:SM}
\end{equation}
with 
\begin{equation}
I_{\gamma\gamma}\equiv I_{\gamma\gamma}(r_f,r_W) = \sum_f Q_f^2 N_{c,
  f} A_{1/2}(r_f) - A_1(r_W)\,,
\label{Igg}
\end{equation}
and 
\begin{align}
A_{1/2}(r_f) &= 2 r_f \left [ 1 + (1 - r_f) f(r_f) \right
]\,, \label{eq:A12} \\
A_1(r_W) &= 2 + 3 r_W \left [ 1 + (2 - r_W) f(r_W) \right ]
\,.  \label{eq:A1}
\end{align}
Here $Q_f$ and $m_f$ are the fermion charge (in the units of proton
charge), and mass, respectively, $N_{c,f}$ is the colour factor for
fermions (3 for quarks, 1 for leptons) and
\begin{equation}
r_f \equiv \frac{4 m_f^2}{M_h^2}\,, \qquad r_W \equiv \frac{4
  M_W^2}{M_h^2} \,.
\end{equation}
The result is of course finite and is governed by a single function
$f(r)$, which reads
\begin{equation}
f(r)=
\begin{cases}
\arcsin^2 \left ( \frac{1}{\sqrt{r}} \right ) \,, \quad r \ge 1 \,,\\
- \frac{1}{4} \left [\log \left ( \frac{1 + \sqrt{1 - r}}{1 - \sqrt{1
       - r}} \right ) - i \pi \right ]^2 \,, \quad r \le 1 \,.
\end{cases}
\label{eq:fr}
\end{equation}
It is useful for order of magnitude calculations to state that
$A_1(r_W) \approx 8.33$, $A_{1/2}(r_t) \approx 1.38$ and
$I_{\gamma\gamma} \approx - 6.56$ with a negligible imaginary part.

The expression given in eq.~\eqref{eq:SM} is \emph{not} exactly the SM
contribution for it is written in terms of SMEFT parameters and not in
terms of measurable quantities like those listed in
eq.~\eqref{expvalues}.
We therefore rewrite eq.~\eqref{eq:SM} in terms of physical quantities
using the expression for $\bar{e}$ from eq.~\eqref{eq:ebar} and $G_F$
from eq.~\eqref{eq:GF} that bring in the new coefficients $C^{\varphi
  W B}$ and $C_{11}^{\varphi l (3)},C_{22}^{\varphi l (3)},
C_{1221}^{ll}$, respectively,
\begin{equation}
\frac{\overline{\Gamma}^{\mathrm{SM}}}{M_W} =
\frac{\alpha_{EM}}{16\pi} \left (\frac{8 G_F}{\sqrt{2}} \right
)^{1/2} I_{\gamma\gamma} \left [ 1 + 2 s c\,
  \frac{v^2}{{\Lambda^2}} C^{\varphi W B} -
  \frac{v^2}{2{\Lambda^2}} ( C_{11}^{\varphi l (3)} +
  C_{22}^{\varphi l (3) } ) + \frac{v^2}{2 {\Lambda^2}}
  C_{1221}^{ll} \right ] \,.
\label{eq:SM:SMEFT}
\end{equation}
Note that the piece before the square brackets on the RHS is the SM
contribution to amplitude [up to a Lorentz factor in
  eq.~\eqref{physamp2}], as it would be calculated in the absence of
any higher order operators.  Inside the square brackets there are
contributions from SMEFT \ie running Wilson coefficients evaluated at
a scale $\mu$.  Hence, the precise determination of the
$\mathcal{R}_{h\to \gamma\gamma}$ in eq.~\eqref{Rhgg0} is
\begin{equation}
\mathcal{R}_{h\to \gamma\gamma}=\frac{\Gamma(\mathrm{SMEFT},h\to
  \gamma\gamma)}{\Gamma(\mathrm{SM},h\to \gamma\gamma)} \equiv 1
+ \delta \mathcal{R}_{h\to \gamma\gamma} \,,
 \label{Rhgg2}
\end{equation}
where the SM decay width reads, in accordance with standard
refs.~\cite{Gunion:1989we, Djouadi:2005gi, Marciano:2011gm}, as
\begin{equation}
\Gamma(\mathrm{SM},h\to \gamma\gamma)=\frac{ G_F\,
  \alpha_{\mathrm{EM}}^2 \, M_h^3}{128 \sqrt{2} \pi^3} \,
|I_{\gamma\gamma}|^2 \,,
\end{equation}
with $I_{\gamma\gamma}$ given in eq.~\eqref{Igg}.  The SMEFT
contributions of eq.~\eqref{eq:SM:SMEFT} are encoded in a part of
$\delta \mathcal{R}_{h\to \gamma\gamma}$ of eq.~\eqref{Rhgg2}, in
terms of measurable quantities $s,c$ and $G_F$, as
\begin{equation}
\delta \mathcal{R}_{h\to \gamma\gamma}^{(1)} \simeq \frac{4 s
  c}{\sqrt{2}} \frac{1}{G_F \Lambda^2} C^{\varphi WB} -
\frac{1}{\sqrt{2}} \frac{1}{G_F \Lambda^2}  ( C_{11}^{\varphi \ell
  (3)} + C_{22}^{\varphi \ell (3) } ) + \frac{1}{\sqrt{2}}
\frac{1}{G_F \Lambda^2}  C_{1221}^{\ell\ell} \,,
\label{eq:R1}
\end{equation}
where $c^2=1 - s^2=M_W^2/M_Z^2$.  Following our EFT expansion
assumption, in obtaining \eq{eq:R1}, corrections of
$\mathcal{O}(1/\Lambda^4)$ have been consistently ignored.

\subsection{$C^{\varphi D}$, $C^{\varphi\Box}$, $C^{\varphi}$}

A direct calculation shows that the contribution from operators
$C^{\varphi \Box}$ and $C^{\varphi D}$ is simply
\begin{equation}
\left ( 1 + {\frac{v^2}{\Lambda^2}} C^{\varphi \Box} -
{\frac{v^2}{4\Lambda^2}} C^{\varphi D} \right ) (i
\mathcal{A}^{\mathrm{SM}}) \equiv Z_h^{ - 1} (i
\mathcal{A}^{\mathrm{SM}}) \,,
\label{cfcd}
\end{equation}
where $Z_h$ is the field redefinition factor for making the kinetic
term of the Higgs field canonical in going from SM to SMEFT (see
eq.(3.5) of ref.~\cite{Dedes:2017zog}) and $i
\mathcal{A}^{\mathrm{SM}}$ is the full SM contribution to $h\to
\gamma\gamma$ amplitude.  There is an explanation for this result
based on the quantization of SMEFT presented in
ref.~\cite{Dedes:2017zog}.  In unitary gauge these operators appear in
Higgs boson vertices ($hWW$ and $hff$) with exactly the same Lorentz
structure as in the corresponding SM vertices.  On the other hand, in
``renormalisable'' gauges these operators appear in a complicated way
e.g., there are contributions from Goldstone bosons $hG^0G^0$ that
have a non-trivial, non-SM Lorentz structure~\cite{Dedes:2017zog} and
eq.~\eqref{cfcd} is not easily seen without performing the actual
calculation.  However, the result should be independent on the gauge
choice as we explicitly confirm.
We can view eq.~\eqref{cfcd} in a different way starting from the SM
amplitude and perform the redefinition $H=Z_h^{-1} h$ on the single
external Higgs boson leg.

As we already mentioned in section~\ref{sec:operators}, the
coefficient $C^\varphi$ does not contribute explicitly to the $h\to
\gamma\gamma$ amplitude in unitary gauge.  Although there are apparent
non-trivial contributions from it to vertices in $R_\xi$-gauges,
once again, gauge invariance implies that the amplitude is explicitly
independent of $C^\varphi$.  Again, we explicitly verify this
situation as well.

In summary, the contribution of operators discussed in this subsection
to the ratio \eqref{Rhgg2} reads trivially, up to $\sim 1/\Lambda^2$
terms, as
\begin{equation}
\delta \mathcal{R}_{h\to \gamma\gamma}^{(2)} \simeq \sqrt{2}
\frac{1}{G_F \Lambda^2} C^{\varphi \Box} - \frac{\sqrt{2}}{4}
\frac{1}{G_F \Lambda^2} C^{\varphi D} \,.
\label{eq:R2}
\end{equation}

\subsection{$C^{e \varphi }$, $C^{u \varphi}$, $C^{d \varphi}$}

The relevant diagrams for these operators contain a fermion
circulating in the loop.  They contribute a $\xi$-independent piece in
the last term of eq.~\eqref{physamp} which takes the form
\begin{equation}
\Gamma^{f\varphi}_{i} = -
\frac{1}{4\pi^2}\frac{\bar{g}^{2}\bar{g}^{\prime 2}}{\bar{g}^{2} +
  \bar{g}^{\prime 2}} N_{c,f} Q^{2}_{f} \frac{v
  m_{f_{i}}}{\sqrt{2} M_{h}^{2}} \left[1 + (1 -
  r_{f_{i}})f(r_{f_{i}})\right]\,.
 \label{eq:r3}
\end{equation}
The contribution runs over all charged fermions $f=e,u,d$ with their
generation flavours denoted as $i=1,2,3$, \ie $u_1=u, u_2=c, u_3=t$
etc.  The electromagnetic charges $Q_f$ and colour factors $N_{c,f}$,
are given in and below eq.~\eqref{eq:charges}.  The function $f(r)$ is
defined in eq.~\eqref{eq:fr}.  Turning all parameters into measurable
ones in eq.~\eqref{eq:r3} we obtain for the $\mathcal{R}_{h\to
  \gamma\gamma}$ ratio of eq.~\eqref{Rhgg2}
\begin{align}
\delta \mathcal{R}_{h\to \gamma\gamma}^{(3)} & \simeq -
\frac{2^{3/4} }{ (G_F M_h^2)^{1/2}} \sum_{f=e,u,d} N_{c,f} Q_f^2 \,
\sum_{i=1}^3 \operatorname{Re}\left [ \frac{A_{1/2}(r_{f_i})}{
    I_{\gamma\gamma} \, r_{f_i}^{1/2}} \right ] \frac{1}{G_F
  \Lambda^2}\, C^{f\varphi}_{ii}\,,
%
\label{eq:R3}
\end{align}
with $A_{1/2}(r)$ being a function defined in eq.~\eqref{eq:A12} {and
  $I_{\gamma\gamma}$ defined in eq.~\eqref{Igg}.}  The function inside
the square parenthesis peaks at the charm mass and as we shall see
below [{\it cf.}~eq.~\eqref{eq:num}] this is the most important
contribution in $\delta \mathcal{R}_{h\to \gamma\gamma}^{(3)}$.

All operators we have examined thus far are of PTG type. These
operators create \emph{only} finite contributions in the $h\to
\gamma\gamma$ amplitude.  On contrary, operators that will be examined
next will need to be renormalised.

\subsection{$C^{\varphi B}$, $C^{\varphi W}$, $C^{\varphi WB}$}

The amplitude in \eq{physamp} contains contributions from $Q_{\varphi
  B}, Q_{\varphi W}$, $Q_{\varphi WB}$ operators\footnote{There is an
  additional contribution from the operator $Q_{\varphi WB}$, arising
  from \eq{eq:SM:SMEFT}, which must be added in the final amplitude,
  {\it cf.}~\eq{eq:num}.}  appearing already at tree level in SMEFT.
These are collected in the first three lines of \eq{physamp}, but
still contain the renormalised vev $v$.  This parameter needs to be
turned into Fermi coupling constant, $G_F$, that is a measurable
quantity with experimental value given in eq.~\eqref{expvalues}.  We
only need the SM one loop corrections to $\Delta r$, which appear
through the expression
\begin{equation}
\frac{\bar{G}_F}{\sqrt{2}} = \frac{1}{2 v^2} \frac{1}{(1 - \Delta
  r)}\,.
\label{eq:GFcor}
\end{equation}
Note that $\Delta r$ is a gauge invariant quantity and its form can be
found in ref.~\cite{Sirlin:1980nh}.  This is consistent with our
remark in section~\ref{sec:renormalization} that the pre-factors of
$C^{\varphi B}, C^{\varphi W}$, $C^{\varphi WB}$ in
eq.~\eqref{physamp} are respectively gauge invariant quantities and
therefore the whole amplitude is gauge invariant.  We then use
eq.~\eqref{eq:GF} to order $1/\Lambda^2$ \ie set $\bar{G}_F \to G_F$
in eq.~\eqref{eq:GFcor} and apply the result in eq.~\eqref{physamp}.
We find that $\Delta r$ nicely cancels out when using an alternative
expression for $\delta v/v$ derived in ref.~\cite{Sirlin:1985ux} in
Feynman gauge $\xi=1$,
\begin{equation}
\frac{\delta v}{v}=\frac{1}{2} \left [
  \frac{A_{WW}(0)}{M_W^2} + \Delta r - \widetilde{E} \right ]_{\xi=1}
\,,
\label{eq:dvv2}
\end{equation}
where the parameter $\widetilde{E}$ is given in
ref.~\cite{Sirlin:1985ux}
\begin{equation}
\widetilde{E}_{\xi=1}=\frac{\alpha_{\mathrm{EM}}}{2 \pi s^2}
\left [ 2 E - 2 \log\frac{M_Z^2}{\mu^2} + \frac{\log c^2}{s^2} \left (
  \frac{7}{4} - 3 s^2 \right ) + 3 \right ]\,.
\end{equation} 
The quantity $A_{WW}(0)$ is presented in ref.~\cite{Marciano:1980pb}
in 't Hooft-Feynman gauge and is recalculated here for
completeness in \eq{eq:AWW0}.  By putting \eqs{eq:GFcor}{eq:dvv2} in
\eq{physamp} we obtain the relevant finite contributions from
operators $Q_{\varphi B},Q_{\varphi W}, Q_{\varphi WB}$, to the
physical amplitude $\mathcal{A}_{h\to \gamma \gamma}$
{\small
\begin{align}
& \frac{c^2\, {C}^{\varphi B}(\mu)}{(\sqrt{2} G_F)^{1/2} {\Lambda^2}}
  \left [ 1 + \Gamma^{\varphi B} - \frac{A_{WW}(0)}{2 M_W^2} +
    \frac{\widetilde{E}}{2} + \frac{1}{2}\Pi_{HH}^\prime (M_h^2) -
    \Pi_{\gamma\gamma}(0) + 2 \tan\theta_W 
    \frac{A_{Z\gamma}(0) + \delta m^2_{Z\gamma}}{M_Z^2}
    \right ]_{\mathrm{finite}} \nonumber \\[2mm]
& + \frac{s^2 \, {C}^{\varphi W}(\mu)}{(\sqrt{2} G_F)^{1/2}
    {\Lambda^2}} \left [ 1 + \Gamma^{\varphi W} - \frac{A_{WW}(0)}{2
      M_W^2} + \frac{\widetilde{E}}{2} + \frac{1}{2}\Pi_{HH}^\prime
    (M_h^2) - \Pi_{\gamma\gamma}(0) - \frac{2}{\tan\theta_W}
    \frac{A_{Z\gamma}(0) + \delta m^2_{Z\gamma}}{M_Z^2} 
    \right ]_{\mathrm{finite}} \nonumber \\[2mm]
& - \frac{s c \, {C}^{\varphi WB}(\mu)}{(\sqrt{2} G_F)^{1/2}
    {\Lambda^2}} \left [ 1 + \Gamma^{\varphi WB} - \frac{A_{WW}(0)}{2
      M_W^2} + \frac{\widetilde{E}}{2} + \frac{1}{2}\Pi_{HH}^\prime
    (M_h^2) - \Pi_{\gamma\gamma}(0) - \frac{2}{\tan 2\theta_W}
    \frac{A_{Z\gamma}(0) + \delta m^2_{Z\gamma}}{M_Z^2}
    \right ]_{\mathrm{finite}}\,.
\label{treesmeft}
\end{align}
}
This expression takes this particular form \emph{only} in $\xi=1$
gauge and replaces the first three lines in eq.~\eqref{physamp}.  It
is important for the reader to notice, that numerically big
corrections from $\Delta r$ have been cancelled out in
eq.~\eqref{treesmeft}.  The quantities $\Gamma^{\varphi V}, V=B,W,WB$
are fairly lengthy and are given in the Appendix~\ref{app:pv} together
with the self-energies, all in general $R_\xi$-gauges.
Nevertheless, following our tactic here, 
we can write down a clear formula for the relevant 
corrections to the ratio $\mathcal{R}^{(4)}_{h\rightarrow \gamma \gamma}$
in eq.\eqref{Rhgg2},  as (recall that $\tan\theta_W = s/c 
= \bar{g}^\prime/\bar{g}$)
{\small
\begin{align}
\delta \mathcal{R}^{(4)}_{h\rightarrow \gamma \gamma} \simeq \frac{8 \pi^{2}}{G_{F}M^{2}_{W} \tan^{2}\theta_{W}}\left[\frac{C^{\varphi B}}{G_{F}\Lambda^{2}} \mathrm{Re}\left(\frac{I_{\varphi B}}{I_{\gamma\gamma}}\right)+\tan^{2}\theta_{W}\frac{C^{\varphi W}}{G_{F}\Lambda^{2}} \mathrm{Re}\left(\frac{I_{\varphi W}}{I_{\gamma\gamma}}\right)-\tan\theta_{W}\frac{C^{\varphi WB}}{G_{F}\Lambda^{2}} \mathrm{Re}
\left(\frac{I_{\varphi WB}}{I_{\gamma\gamma}}\right)\right]_{\mathrm{finite}},
\label{eq:R4}
\end{align}
}
where $I_{\varphi B},\, I_{\varphi W},\, I_{\varphi WB}$  represent the expressions in corresponding squared brackets of eq.~\eqref{treesmeft}.

As we already mentioned in the discussion below \eq{eq:dee}, the
photon self-energy, $\Pi_{\gamma\gamma}(0)$, contains hadronic
contributions from five light quarks \ie all quarks but the top quark.
Therefore, for the related part,
$\Pi_{\gamma\gamma}^{\mathrm{had}}(0)$, the perturbative formula
\eqref{app:Pgg} is not reliable.  We use instead,
\begin{equation}
\Pi_{\gamma\gamma}^{\mathrm{had}}(0) = - \Delta
\alpha^{(5)}_{\mathrm{had}}(M_Z^2) +
\Pi_{\gamma\gamma}^{\mathrm{had}}(M_Z^2)\,,
\label{eq:pgghad}
\end{equation}
where now, thanks to asymptotic freedom,
$\Pi_{\gamma\gamma}^{\mathrm{had}}(M_Z^2)$ is a reliable perturbative
one-loop calculation for the light quark contributions (see
\eqref{app:Pggh}) while $\Delta
\alpha^{(5)}_{\mathrm{had}}(M_Z^2)=\Pi_{\gamma\gamma}^{\mathrm{had}}(M_Z^2)
- \Pi_{\gamma\gamma}^{\mathrm{had}}(0)$ is finite and is computed via
a dispersion relation that involves experimental data for the ratio
$\sigma (e^+ e^- \to \mathrm{hadrons})/\sigma (e^+ e^- \to \mu^+ \mu^-
)$.  A recent analysis~\cite{Patrignani:2016xqp} gives $\Delta
\alpha^{(5)}_{\mathrm{had}}(M_Z^2) = 0.02764 \pm 0.00013$.

The form for $\delta \mathcal{R}_{h\to \gamma\gamma}^{(4)}$ in 
eq.~\eqref{eq:R4}  is given
semi-analytically below [{\it cf.}~eq.~\eqref{eq:num}].  Since these
corrections appear at tree level in SMEFT they are generically the
biggest ones from all operators involved in $h\to \gamma\gamma$
amplitude.

\subsection{$C^{W}$}

The contribution from $W$-loops gives rise to terms proportional to
$C^W$ in eq.~\eqref{physamp}.  The relevant expression is
$\xi$-independent, and is written as
\begin{eqnarray}
\Gamma^{W} = \frac{3}{16\pi^2} \frac{\bar{g}^3 \bar{g}^{\prime
    2}}{(\bar{g}^2 + \bar{g}^{\prime 2})} [3 E + B]\,,
\end{eqnarray}  
where $E$ is the infinite piece [see eq.~\eqref{eq:e}] formed as usual
in dimensional regularisation, of course removed from
eq.~\eqref{physamp}.  The integral function $B$ is
\begin{equation}
B \equiv B(r_W)=2-r_{W} f(r_{W}) + 2 J_{2}(r_{W}) - 3
\log\frac{M_{W}^{2}}{\mu^{2}} \,,
\label{eq:B}
\end{equation}
where the functions $f(r), J_{2}(r)$ are given in eqs.~\eqref{eq:fr}
and \eqref{eq:I2}, respectively, and $\mu$ is the renormalisation
scale.  The contribution from the operator $Q_W$ in the ratio
\eqref{Rhgg2} is
\begin{equation}
\delta \mathcal{R}^{(5)}_{h\rightarrow \gamma \gamma}
\simeq 24\: \sqrt{\frac{G_{F}M^{2}_{W}}{\sqrt{2}}} \: 
\mathrm{Re} \: \left[\frac{B(r_{W})}{I_{\gamma\gamma}}\right] \: \frac{1}{G_{F}\Lambda^{2}}\: C^{W}\;,
  \label{eq:R5}
\end{equation}
with $I_{\gamma\gamma}$ defined in eq.~\eqref{Igg}.

\subsection{$C^{eB}$, $C^{eW}$, $C^{uB}$, $C^{uW}$, $C^{dB}$, $C^{dW}$}

These are again contributions from operators affecting fermion loops
and, as such, they are $\xi$-independent.  They are, however, infinite
since they involve dipole operators (as one can easily see from
ref.~\cite{Dedes:2017zog} there is an extra momentum in the numerator
of their corresponding Feynman rules expressions).  We obtain the
following contribution in the last term of eq.~\eqref{physamp}:
\begin{align}
\Gamma^{fB}_i &= \frac{1}{4\pi^2} 
\frac{\bar{g}^{2}\bar{g}'}{\bar{g}^{2} + \bar{g}^{\prime 2}} \,
N_{c,f} Q_{f} \, \frac{m_{f_{i}}}{\sqrt{2} v} \left[2E
  +D(r_{f_i})\right]\,, \nonumber \\ \Gamma^{fW}_i &= 2 T_f^3\,
\frac{\bar{g}^\prime}{\bar{g}}\, \Gamma^{fB}_i\,,
\end{align}
where  the function $D(r_{f_i})$ is defined as
\begin{equation}
D(r_{f_i})\equiv -2\log\frac{m_{f_i}^2}{\mu^2} +1-r_{f_i} f(r_{f_i}) +
J_2(r_{f_i}) \,.
\label{eq:DD}
\end{equation}
Here again $f$ stands for a fermion type, $f=e,u,d$, and $i=1,2,
3$ runs over its flavour eigenstates.
The relevant contribution from the operators $Q_{fB}$ and $Q_{fW}$ to
the ratio $\mathcal{R}_{h\to \gamma\gamma}$ of eq.~\eqref{Rhgg2} is
\begin{equation}
\delta \mathcal{R}_{h\to \gamma\gamma}^{(6)} \simeq   \frac{2
    M_h}{M_W \tan\theta_W} \sum_{f=e,u,d} N_{c,f} Q_f \sum_{i=1}^3
  \operatorname{Re} \left [ \frac{r_{f_i}^{1/2}D(r_{f_i})}{
      I_{\gamma\gamma} } \right ]\, \frac{1}{G_F \Lambda^2}
  (C^{fB}_{ii} + 2 T_f^3 \tan\theta_W C_{ii}^{fW}) \,.
\label{eq:R6}
\end{equation} 
Functions $I_{\gamma\gamma}, f(r)$ and $J_2(r)$ are defined in
eqs.~\eqref{Igg}, \eqref{eq:fr} and \eqref{eq:I2}, respectively.

The expression $\delta \mathcal{R}_{h\to \gamma\gamma}^{(6)}$ in
eq.~\eqref{eq:R6} has few interesting features. 
It is proportional to the mass of the
fermion circulated in the loop and also proportional to
$\mathcal{O}(1)$ loop functions ratio.  Comparing $\delta
\mathcal{R}_{h\to \gamma\gamma}^{(6)}$, which arises from LG
operators, with, for example, $\delta \mathcal{R}_{h\to
  \gamma\gamma}^{(3)}$ of eq.~\eqref{eq:R3} which arises from PTG
operators and recall Table~\ref{tab:2}, we see that there is a huge
enhancement of the former by a factor of $\mathcal{O}(10)$ in
particular for the top-quark.
Hence, for the top quark in the loop and for $\mu=M_W$, this is the biggest correction
from all one-loop contributions in SMEFT as we shall see shortly in
section~\ref{sec:results}.

\section{Results}
\label{sec:results}

\subsection{Semi-numerical expression for the ratio
  $\mathcal{R}_{h\to \gamma\gamma}$
}

In this section, we sum all contributions to $\mathcal{R}_{h\to
  \gamma\gamma}$ found in section~\ref{sec:amplitude}, leaving as
unknowns, the renormalisation group running Wilson coefficients,
$C=C(\mu)$, the renormalisation scale $\mu$ divided by the $W$-boson
mass and the energy scale $\Lambda$.
Everything we have discussed so far is within the perturbative
renormalisation framework explained in
section~\ref{sec:renormalization}.  For EFT expansion to be valid,
this means that the maximum value of a generic coefficient,
$C/\Lambda^2$, is at most ${\cal O}(1)$.  Experimentally, it is
suggested from \eq{Rhgg} that the corrections to $\delta
\mathcal{R}_{h\to \gamma\gamma}$ should be at most 15\%.
Being conservative, and in order to display all ``important"
contributions from operators in $\delta \mathcal{R}_{h\to
  \gamma\gamma}$, we present below semi-numerical results for $\delta
\mathcal{R}_{h\to \gamma\gamma}$ that are up to $1\% \times
C/\Lambda^2$.

With the energy scale \emph {$\Lambda$ written in TeV units,} we
obtain (in Warsaw basis)\footnote{Unlike refs.~\cite{Hartmann:2015aia,
    Hartmann:2015oia} we have made no rescaling of Wilson coefficients
  with gauge couplings.  Of course, the coefficients-$C^{fB,fW}$ are
  the rotated coefficients in the quark or lepton mass basis adopted
  in ref.~\cite{Dedes:2017zog} as already noted in
  section~\ref{sec:operators}.}:
\begin{align}
    \delta \mathcal{R}_{h\to \gamma\gamma} &=
\sum_{i=1}^6 \delta \mathcal{R}_{h\to \gamma\gamma}^{(i)}
\simeq 0.06 \left (
\frac{C^{\ell\ell}_{1221} - C^{\varphi \ell (3)}_{11} - C^{ \varphi
    \ell (3)}_{22} }{\Lambda^2} \right ) + 0.12 \left (
\frac{C^{\varphi \Box} - \frac{1}{4} C^{\varphi D} }{\Lambda^2} \right
) \nonumber \\&\quad
- 0.01 \left( \frac{C^{e\varphi}_{22} + 4 C^{e\varphi}_{33} + 5
   C^{u\varphi }_{22} + 2 C^{d\varphi }_{33} - 3 C^{u \varphi
  }_{33}}{\Lambda^2} \right ) \nonumber \\&\quad
- \left[48.04 - 1.07 \log
  \frac{\mu^{2}}{M^{2}_{W}}\right]\frac{C^{\varphi
    B}}{\Lambda^{2}}
- \left[14.29 - 0.12 \log
  \frac{\mu^{2}}{M^{2}_{W}}\right]\frac{C^{\varphi
  W}}{\Lambda^{2}} \nonumber \\&\quad
+ \left[26.62 - 0.52 \log
  \frac{\mu^{2}}{M^{2}_{W}}\right]\frac{C^{\varphi
  WB}}{\Lambda^{2}} \nonumber \\&\quad
+ \left [ 0.16 - 0.22 \log
    \frac{\mu^{2}}{M^{2}_{W}}\right]\frac{C^{W}}{\Lambda^{2}}
    \nonumber \\&\quad
+ \left [ 2.11 - 0.84 \log
    \frac{\mu^{2}}{M_W^2}\right]\frac{C^{uB}_{33}}{\Lambda^{2}}
  + \left [ 1.13 - 0.45 \log
    \frac{\mu^{2}}{M_W^2}\right]\frac{C^{uW}_{33}}{\Lambda^{2}}
    \nonumber \\&\quad
- \left [ 0.03 + 0.01 \log
    \frac{\mu^{2}}{M_W^2}\right]\frac{C^{uB}_{22}}{\Lambda^{2}}
  - \left [ 0.01 + 0.00 \log
    \frac{\mu^{2}}{M_W^2}\right]\frac{C^{uW}_{22}}{\Lambda^{2}}
    \nonumber \\&\quad
+ \left [ 0.03 + 0.01 \log
    \frac{\mu^{2}}{M_W^2}\right]\frac{C^{dB}_{33}}{\Lambda^{2}}
  - \left [ 0.02 + 0.01 \log
    \frac{\mu^{2}}{M_W^2}\right]\frac{C^{dW}_{33}}{\Lambda^{2}}
    \nonumber \\&\quad
+ \left [ 0.02 + 0.00 \log
    \frac{\mu^{2}}{M_W^2}\right]\frac{C^{eB}_{33}}{\Lambda^{2}}
  - \left [ 0.01 + 0.00 \log
    \frac{\mu^{2}}{M_W^2}\right]\frac{C^{eW}_{33}}{\Lambda^{2}}
  + \ldots \,,
\label{eq:num}
\end{align}
where the ellipses denote contributions from the operators $Q$ in
Table~\ref{tab:no4ferm} that are less than $1\% \times C/\Lambda^2$.
Terms in the first three parentheses arise from finite loop
contributions, $\delta \mathcal{R}_{h\to \gamma\gamma}^{(1,2,3)}$ in
eqs.~\eqref{eq:R1}, \eqref{eq:R2} and \eqref{eq:R3}, while all the
rest arise from ``infinite'' diagrams; for these the renormalisation
scale $\mu$ appears explicitly.  All coefficients are running
quantities, $C=C(\mu)$, and $\delta \mathcal{R}_{h\to \gamma\gamma}$
should be RGE invariant up to one-loop and up to $1/\Lambda^2$
expansion terms.  This can be checked numerically already from the
explicit $\mu$-dependence in \eq{eq:num} and the $\beta$-functions for
the $C$-coefficients calculated in refs.~\cite{Jenkins:2013zja,
  Jenkins:2013wua, Alonso:2013hga}.\footnote{For this purpose, one can
  use the numerical codes of \Refs{Celis:2017hod,Aebischer:2018bkb} or
  can exploit analytic techniques appeared recently in
  \Ref{Buras:2018gto}.}
Furthermore, we remark that in eq.~\eqref{eq:num} and for $\mu=1$ TeV,
the logarithmic parts are of the same order of magnitude as the
finite, constant, parts.  Interestingly, for the coefficients in the
last three lines of eq.~\eqref{eq:num}, the two parts constructively
interfere, while for the rest of coefficients they partially cancel.

At the end of the day, only five operators in eq.~\eqref{eq:num} can
be bounded by the LHC experimental measurement \eqref{Rhgg} of the
ratio $R_{h\to \gamma\gamma}$.  Taking $\mu=M_W$, we find
\begin{eqnarray}
\frac{|C^{\varphi B}|}{\Lambda^2} \lesssim \frac{0.003}{(1~\mathrm{TeV})^2}\,, & \qquad &
\frac{|C^{\varphi W}|}{\Lambda^2} \lesssim \frac{0.011}{(1~\mathrm{TeV})^2}\,, \qquad
\frac{|C^{\varphi WB}|}{\Lambda^2} \lesssim \frac{0.006}{(1~\mathrm{TeV})^2}\,, \nonumber \\[2mm]
\frac{|C^{uB}_{33}|}{\Lambda^2} \lesssim \frac{0.071}{(1~\mathrm{TeV})^2} \,, &\qquad &
\frac{|C^{uW}_{33}|}{\Lambda^2} \lesssim \frac{0.133}{(1~\mathrm{TeV})^2} \,.  \label{eq:bounds}
\end{eqnarray}
All bounded coefficients above are associated with LG operators in
Table~\ref{tab:2} in a perturbative decoupled
UV-theory. \Eq{eq:bounds} seems to be consistent with this observation
and $\Lambda \approx 1$ TeV.
On the other hand, assuming $|C^{\varphi V}|\; (|C^{uB,uW}_{33}|)
\simeq 1$ we obtain $\Lambda \gtrsim 10\; (3) $ TeV, outside but close
to the near-future LHC region.  Other operators in eq.~\eqref{eq:num}
{may} contribute at most 15\% only when $C=1$ and $\Lambda = 1$ TeV
{so their effects are less likely to be observed} at present in LHC
searches for the $h\to \gamma\gamma$ process.

Operators $Q_{\varphi B}$, $Q_{\varphi W}$ and $Q_{\varphi WB}$
contribute already at tree level in SMEFT and this explains the
large value of their coefficients in eq.~\eqref{eq:num}.  As our
calculation shows, taking also into account one-loop corrections,
modify their respective tree level contributions to the ratio $\delta
R_{h\to \gamma\gamma}$ by 1.3\% for $C^{\varphi B}$, by 7.5\% for
$C^{\varphi WB}$ and by 8.7\% for $C^{\varphi W}$ at the
renormalisation scale $\mu=M_W$, in agreement with the commonly
expected magnitude of the SM-like electroweak one-loop corrections.
What is surprising however, is the large loop contribution of dipole
operators $Q_{uB,uW}^{33}$.  This is basically due to the largeness of
the top-quark mass and other features already noted in the discussion
below eq.~\eqref{eq:DD}.

\subsection{Other constraints}

In the section above, we found that the dominant coefficients in
$\mathcal{R}_{h\to \gamma\gamma}$ are those given in
eq.~\eqref{eq:bounds}.  These coefficients maybe also bounded by
observables other than $h\to \gamma\gamma$.  It has been noted in
refs.~\cite{Grinstein:1991cd, Peskin:1991sw} that the coefficient
$C^{\varphi W B}$ contributes directly to the electroweak
$S$-parameter, one of the parameters that fits $Z$-pole observables.
Its contribution reads
\begin{equation}
\frac{C^{\varphi WB}}{\Lambda^2} = \frac{G_F\, \alpha_{\mathrm{EM}}
}{2 \sqrt{2} s c} \Delta S \,.
\end{equation}
With $\Delta S \in [-0.06,0.07]$ \cite{Ellis:2018gqa} we obtain
$\frac{|C^{\varphi WB}|}{\Lambda^2} \lesssim 0.005~\mathrm{TeV}^{-2}$
which is of the same order of magnitude as the upper bound we find
here in eq.~\eqref{eq:bounds} from $h\to \gamma\gamma$ measurement.
The coefficients $C^{\varphi W}$ and $C^{\varphi B}$ are constrained
by LHC Higgs data (giving upper limits on deviations from the SM
predictions) or electroweak fits to EW observables. The respective
bounds, as they read from refs.~\cite{Falkowski:2014tna,
  Ellis:2018gqa}, are also about the same order of magnitude as in
eq.~\eqref{eq:bounds}.

The other two operators in eq.~\eqref{eq:bounds}, $Q_{uB}^{33}$ and
$Q_{uW}^{33}$, are constrained from the $\bar{t}tZ$ production and the
latter also by the single top production measurements at LHC.  Bounds
quoted in ref.~\cite{Buckley:2015lku} are $|C^{uB}_{33}|/\Lambda^2
\lesssim 7.1~\mathrm{TeV}^{-2}$ and $|C^{uW}_{33}|/\Lambda^2 \lesssim
2.5~\mathrm{TeV}^{-2}$.  Here, bounds from $h\to \gamma\gamma$ derived
in eq.~\eqref{eq:bounds} are \emph{more than an order of magnitude}
\emph{stronger}.

Restrictions to all other coefficients appeared in eq.~\eqref{eq:num}
can be found in various articles in the literature.  For example,
following ref.~\cite{Ellis:2018gqa}, $Q_{\varphi D}$ contributes to
the $T$-electroweak parameter and the corresponding bound is,
$|C^{\varphi D}|/\Lambda^2 \lesssim 0.03~\mathrm{TeV}^{-2}$.  This
makes its contribution in $h\to \gamma\gamma$ negligible.  However,
the coefficients $C^{\varphi \Box}$ and $C^{W}$ are not really
constrained by fitting the LHC Higgs data.  It is obvious from
eq.~\eqref{eq:num} that these two coefficients can give ${\cal
  O}(10)$\% contributions to $\mathcal{R}_{h\to \gamma\gamma}$ only
when one is in the vicinity of EFT validity.

\subsection{Comparison with  literature}
As we mentioned in the introduction, the calculation for \hgg in SMEFT
was first performed several years ago in
\Refs{Hartmann:2015aia,Hartmann:2015oia} and to our knowledge these
are the only complete studies prior to ours here. Our check shows that there
are two, numerically important differences. First, all corresponding
$\delta \mathcal{R}_{h\to \gamma\gamma}$ in \Ref{Hartmann:2015aia} are smaller by exactly a factor of four. We think
that this is due to a mistake in eq.~(26) of
\Ref{Hartmann:2015aia}[arXiv v3].  Second, our \eq{eq:R3} is
\emph{not} in agreement with the corresponding expression of
ref.~\cite{Hartmann:2015aia}. We believe there is a Yukawa coupling
missing for each generation and flavour in the corresponding
expression of \Ref{Hartmann:2015aia}.
Up to the aforementioned differences, we found agreement
with $\delta \mathcal{R}_{h\to \gamma\gamma}^{(1,2,3,5,6)}$.
As far as $\delta \mathcal{R}_{h\to \gamma\gamma}^{(4)}$ is concerned,
a direct comparison of our formulae in \eq{treesmeft} with the
corresponding one in \Ref{Hartmann:2015oia} is very difficult.
Checking individually quantities appearing in both works, for example,
$\delta v/v$ or $\Pi_{HH}'$, is meaningless since the calculations in
\Refs{Hartmann:2015aia,Hartmann:2015oia} were performed in background
field gauges while ours in linear $R_\xi$-gauges.
Comparing numerically the correction, $\delta \mathcal{R}_{h\to
  \gamma\gamma}^{(4)}$, appearing in our \eq{eq:num} with a
corresponding ratio based on refs.~\cite{Hartmann:2015aia,Hartmann:2015oia}, 
we find, upon
fixing the factor of four mentioned above, a maximal difference of 5\%
for $\mu=M_W$, originating from what multiplies the coefficient
$C^{\varphi B}$.

\section{Conclusions}
\label{sec:conclusions}

In our analysis we have calculated the one-loop decay width of the
$h\to \gamma\gamma$ process in the SM extended by all CP-conserving
gauge invariant operators up to dimension-6 in Warsaw basis. We
performed the calculations using the general $R_\xi$-gauges and a
hybrid renormalisation scheme, where we assumed the on-shell
conditions for the SM parameters and $\overline{\mathrm{MS}}$
subtraction for the running Wilson coefficients of the higher order
operators.  We explicitly checked the gauge $\xi$-parameter
cancellation, which provides the very strict test of correctness of
our calculations. In addition, we also explicitly proven that at the
one-loop and $1/\Lambda^2$ order, the calculated amplitude is
independent of the renormalisation scale $\mu$. Our work is
complementary to previous analyses~\cite{Hartmann:2015aia,
  Hartmann:2015oia} of this process using the Background Field Method
and comparisons of our results with theirs were made whenever
possible. Our master formula for the $S$-matrix amplitude is given by
\eqs{physamp2}{physamp}.

We give a complete set of analytical formulae for all classes of SM
and SMEFT contributions to $h\to \gamma\gamma$ decay rate, normalised
to the SM result as in published LHC searches [see \eq{Rhgg2}].  
We also present them in a form of simple and compact semi-analytical
expressions 
depending only on running Wilson coefficients and
renormalisation scale $\mu$.  
\Eq{eq:num} summarises all dominant
contributions.  Such formula can be readily used as additional
constraint in experimental or theoretical analyses considering other
observables in SMEFT.

We show that numerically largest corrections to the SM prediction can
arise from $Q_{\varphi B}$, $Q_{\varphi W}$ and $Q_{\varphi WB}$
operators, contributing already at the tree level, and from
$Q_{uB}^{33}$, $Q_{uW}^{33}$ operators arising at the loop level.
Only Wilson coefficients of these operators can be meaningfully
constrained using the current precision of the LHC measurements for
the $h\to \gamma\gamma$ decay width. In some cases, like $C^{uB}_{33}$
and $C^{uW}_{33}$, such constraints are already stronger than those
from other measurements, in this case for instance from top-quark
LHC-physics.


It would be useful to connect our main outcome, the expression 
eq.~\eqref{eq:num}, with a particular UV-model. One may follow ref.~\cite{deBlas:2017xtg}
in integrating out heavy fields, which 
under reasonable assumptions but limited to perturbative decoupling at
tree-level,  results in a subset of operators arranged in Table~\ref{tab:no4ferm}. Interestingly, one can arrange a finite number of heavy fields with renormalizable (or not) interactions that affect both PTG and LG operators in Table~\ref{tab:2}. Another possibility may be a direct model like the one of ref.~\cite{Manohar:2013rga} where the  operators,
$Q_{\varphi B}, Q_{\varphi W}$ and $Q_{\varphi WB}$, 
are generated. In general however, it is quite difficult, if possible in any way, to find a 
model with appreciable,  $\mathcal{O}(1)$, coefficients for these operators. 
 Possibly, some examples will be found in the future.

A general look of our SMEFT calculational framework does not differ
from common frameworks calculating electroweak one-loop corrections,
like in the renormalisable SM for example.
Our work can easily be automatised although we performed as many
manual calculations we could for comparisons and cross checks.  For
example, one can use the SMEFT Feynman rules, given also in a {\em
  Mathematica} code, from ref.~\cite{Dedes:2017zog}, and existed codes
to calculate Feynman diagrams, employ a ``traditional" renormalisation
prescription from 80's described also here and, checking gauge
invariance at every step, present a concise form of an amplitude in a
useful semi-numeric form, as in \eq{eq:num}.  It is worth for pursuing
this SMEFT framework further.

\subsection*{Acknowledgements}

%
The work of MP is supported in part by the National Science Centre,
Poland, under research grant
DEC-2015/19/B/ST2/02848.
The work of JR is supported in part by the National Science Centre,
Poland, under research grants DEC-2014/15/B/ST2/02157 and
DEC-2016/23/G/ST2/04301.
KS would like to thank the Greek State Scholarships Foundation (IKY)
for full financial support through the Operational Programme ``Human
Resources Development, Education and Lifelong Learning, 2014-2020''.
AD and KS would like to thank University of Warsaw for hospitality.
JR would also like to thank to University of Ioannina and to CERN for
hospitality during his visits there.
AD, JR, and KS would also like to thank M. Misiak for enlightening
discussions on renormalisation, anomalies and evanescent operators in
SMEFT. AD would like to thank  C. Foudas for bringing to our attention
ref.\cite{Sirunyan:2018ouh}.

\newpage
\appendix

\section{SMEFT amplitudes and SM self-energies in $R_{\xi}$-gauges}
\label{app:pv}

We append here the one-loop corrections in general renormalisable
gauges for the three-point 1PI functions, $\Gamma^{\varphi B}$, $\Gamma^{\varphi W}$ and
$\Gamma^{\varphi WB}$, as well as for the SM vector boson self-energies that
are needed for \eqs{physamp}{treesmeft}. The first, $\xi$-independent,
terms of the equations below refer always to a part in unitary gauge.
The {\em Mathematica} package
\texttt{FeynCalc}~\cite{Mertig:1990an, Shtabovenko:2016sxi} was used
for most of our Feynman diagram calculations. To bring Feynman
integrals into analytic forms we used the {\em Mathematica} package
\texttt{Package-X}~\cite{Patel:2015tea, Patel:2016fam}.  In what
follows, we use the mass-ratios
\begin{equation}
    r_{X}\equiv\frac{4 M^{2}_{X}}{M^{2}_{h}}\qquad\text{and}\qquad r_{XY}\equiv\frac{4
  M_{X}^{2}}{M_{Y}^{2}}\,.
\end{equation}
For the SMEFT one-loop corrections
we have
\begin{align}
\label{CB}
\Gamma^{\varphi B}=\frac{-\lambda}{32\pi^{2}}\bigg\{&
3\left(E+2-\frac{\pi}{\sqrt{3}}-\log\frac{M_{h}^{2}}{\mu^{2}}\right)
+2\left(E+2-\log\frac{M_{W}^{2}}{\mu^{2}}-\log\xi_{W}+
J_{2}(\xi_{W} r_{W})\right) \nonumber\\&
+E+2-\log\frac{M_{Z}^{2}}{\mu^{2}}-\log\xi_{Z}+
J_{2}(\xi_{Z} r_{Z}) \bigg\} \,,
\end{align}
\begin{align}
\label{CW}
\Gamma^{\varphi W}=\frac{-1}{32\pi^{2}}\bigg\{&
3\lambda\left(E+2-\frac{\pi}{\sqrt{3}}-\log\frac{M_{h}^{2}}{\mu^{2}}\right)
+\bar{g}^{2} \left[6r_{W} \left(1-r_{W} f(r_{W})\right)
-16(1-r_{W})f(r_{W})\right] \nonumber\\&
+2\left(\lambda-\bar{g}^{2}(\xi_{W}+3)\right)
\left(E-\log\frac{M_{W}^{2}}{\mu^{2}}-\log\xi_{W}\right) \nonumber \\&
+4\lambda-\bar{g}^{2}(\xi_{W}+5)
+\frac{6\bar{g}^{2}}{\xi_{W}-1}\log\xi_{W}
+2\lambda J_{2}(\xi_{W} r_{W}) \nonumber\\&
+\lambda\left(E+2-\log\frac{M_{Z}^{2}}{\mu^{2}}-\log\xi_{Z}
+J_{2}(\xi_{Z} r_{Z})\right) \bigg\} \,,
\end{align}
\begin{align}
\label{CWB}
\Gamma^{\varphi W B}=\frac{-1}{32\pi^{2}}\bigg\{&
-\lambda \left(E+2+\sqrt{3}\pi -\log\frac{M_{W}^{2}}{\mu^{2}}\right)
+6\bar{g}^{2} \left(E-\log\frac{M_{W}^{2}}{\mu^{2}}\right)
+\frac{2\bar{g}^{2}\bar{g}^{\prime 2} \left(3 \bar{g}^{2} + 2 \lambda
\right)}{\lambda(\bar{g}^{2} + \bar{g}^{\prime 2})} \nonumber\\&
-3\lambda \log\frac{M_{h}^{2}}{M_W^{2}}
-\frac{2\bar{g}^{2}(3\bar{g}^{2}\bar{g}^{\prime 2}+2\lambda\bar{g}^{2}-4\lambda\bar{g}^{\prime 2})}
{\lambda (\bar{g}^{2} + \bar{g}^{\prime 2})}
r_{W} f(r_{W})+2(\bar{g}^{2}-2\lambda)J_{2}(r_{W}) \nonumber\\&
-\frac{16}{M^{2}_{h}}
\frac{\bar{g}^{2} \bar{g}^{\prime 2}}{\bar{g}^{2}+\bar{g}^{\prime 2}}
\sum_{f}m^{2}_{f} Q^{2}_{f} N_{c,f}
\left[1+(1-r_{f})f(r_{f})\right] \nonumber\\&
+\lambda \left(E+2-\log\frac{M_{Z}^{2}}{\mu^{2}}-\log\xi_{Z}
+J_{2}(\xi_{Z} r_{Z})\right) \nonumber\\&
+\left(2\lambda-\bar{g}^{2} \left(\xi_{W}+3\right) \right)
\left(E - \log\frac{M_{W}^{2}}{\mu^{2}}-\log\xi_{W}\right)
 \nonumber\\&
+4\lambda-\frac{\bar{g}^{2}}{2}(\xi_{W}+5)+\frac{3\bar{g}^{2}}{\xi_{W}-1} \log\xi_{W}
+2\lambda J_{2}(\xi_{W} r_{W}) \bigg\}\,.
\end{align}

The SM self-energies are presented (to our knowledge for the first
time) also in ref.~\cite{Degrassi:1992ff}, for general renormalisable
gauges, and in ref.~\cite{Marciano:1980pb} for $\xi=1$.  We have
recalculated them here for consistency. The results are:
\begin{align}
\label{app:Pgg}
\Pi_{\gamma \gamma}(0)=& -\frac{1}{48\pi^{2}}
\frac{\bar{g}^{2} \bar{g}^{\prime 2}}{\bar{g}^{2}+\bar{g}^{\prime 2}}
\Bigg[ 21\left(E-\log\frac{M_{W}^{2}}{\mu^2}\right)+2
-4\sum_{f}N_{c,f} Q_{f}^{2}
\bigg(E-\log\frac{m_{f}^{2}}{\mu^2}\bigg) \Bigg] \nonumber\\&
+\frac{1}{32\pi^{2}}
\frac{\bar{g}^{2}\bar{g}^{\prime 2}}{\bar{g}^{2}+\bar{g}^{\prime 2}}
\left[ 2\left(\xi_{W}+3\right)
\left(E-\log\frac{M_{W}^{2}}{\mu^{2}}\right) +\xi_{W}+5
+\frac{2\xi_{W}\left(\xi_{W}+2\right)}{1-\xi_{W}} \log\xi_{W}\right] \,,
\end{align}
\begin{align}
A_{Z \gamma}(0)=
\frac{\bar{g}^{3}\bar{g}^{\prime}v^{2}}{(16\pi)^{2}}
\left[ 2\left(\xi_{W}+3\right)
\left(E-\log\frac{M_{W}^{2}}{\mu^{2}}\right)
+\xi_{W}+5 +\frac{2\xi_{W}\left(\xi_{W}+2\right)}{1-\xi_{W}}
  \log \xi_{W} \right] \,,
\end{align}
\begin{align}
A_{ZZ}(M^{2}_{Z})&=\frac{v^{2}}{768\pi^{2}}\Bigg\{\left(59
\bar{g}^{4} - 36 \bar{g}^{2}\bar{g}^{\prime 2} -
11\bar{g}^{\prime 4}\right)E \nonumber \\&\quad
+\frac{2\left(278\bar{g}^{6} +
  29\bar{g}^{4}\bar{g}^{\prime 2} - 140\bar{g}^{2}\bar{g}^{\prime
    4} - 24\lambda^{2}(\bar{g}^{2} + \bar{g}^{\prime 2}) +
  36\lambda (\bar{g}^{2} + \bar{g}^{\prime 2})^{2} -
  35\bar{g}^{\prime 6}\right)}{3(\bar{g}^{2} + \bar{g}^{\prime 2})} 
  \nonumber \\&\quad
 + \lambda\left(\frac{32\lambda^{2}}{\bar{g}^{2} +
  \bar{g}^{\prime 2}} - 48\lambda + 36(\bar{g}^{2} +
\bar{g}^{\prime 2})\right)\log\frac{M^{2}_{h}}{\mu^{2}}
\nonumber \\&\quad
 + 2\left(\frac{ - 16\lambda^{3}}{\bar{g}^{2} + \bar{g}^{\prime 2}}
+ 24\lambda^{2} - 18\lambda (\bar{g}^{2} + \bar{g}^{\prime 2}) +
5(\bar{g}^{2} + \bar{g}^{\prime
  2})^{2}\right)\log\frac{M^{2}_{Z}}{\mu^{2}} \nonumber \\&\quad
 + \left( - 69\bar{g}^{4} + 16\bar{g}^{2}\bar{g}^{\prime 2} +
\bar{g}^{\prime 4}\right)\log\frac{M^{2}_{W}}{\mu^{2}}
\nonumber \\&\quad
 + \frac{(3\bar{g}^{2} - \bar{g}^{\prime 2})(33\bar{g}^{4} +
  22\bar{g}^{2}\bar{g}^{\prime 2} + \bar{g}^{\prime
    4})}{\bar{g}^{2} + \bar{g}^{\prime 2}} J_{2}(r_{WZ}) \nonumber \\&\quad
 - 16\left[4\lambda^{2} - 4\lambda (\bar{g}^{2} + \bar{g}^{\prime
  2}) + 3 (\bar{g}^{2} + \bar{g}^{\prime 2})^{2}\right] J_{1}(r_{Z})
  \nonumber \\&\quad
+16(\bar{g}^{2}+\bar{g}^{\prime
2})^{2} \sum_{f}N_{c,f}\nonumber\\&\qquad
   \times \bigg\{ g^{2}_{A,f}\bigg[\left(\frac{3}{2}
  r_{fZ} - 1 \right)\bigg(E -
  \log{\frac{m^{2}_{f}}{\mu^{2}}}\bigg)+ 2 r_{fZ} 
-\frac{5}{3} + \left( r_{fZ} - 1 \right) J_{2}(r_{fZ})\bigg]
\nonumber\\&\qquad\quad
-g^{2}_{V,f}\bigg[ E - \log{\frac{m^{2}_{f}}{\mu^{2}}} + r_{fZ}
    +\frac{5}{3} + \left(\frac{1}{2}r_{fZ} + 1\right)
  J_{2}(r_{fZ})\bigg]\bigg\}
\nonumber\\&\quad
-6\xi_{W}\bar{g}^{2}(\bar{g}^{2} + \bar{g}^{\prime 2})\left(E + 1 -
\log{\xi_{W}} - \log\frac{M^{2}_{W}}{\mu^{2}}\right)
\nonumber \\&\quad
 - 3\xi_{Z}(\bar{g}^{2} + \bar{g}^{\prime 2})^{2}\left(E + 1 -
\log{\xi_{Z}} -
\log\frac{M^{2}_{Z}}{\mu^{2}} \right)\Bigg\} \,,
\end{align}
where the axial-vector and vector couplings are defined as
$g_{A,f}=\frac{1}{2} T^{3}_{f}$ and $g_{V,f}=\frac{1}{2} T^{3}_{f} -
\sin^{2}\theta_{w} Q_{f}$, respectively.  The neutrino term in
$A_{ZZ}(M_{Z}^{2})$ is contained in the fermionic part, and can
readily be obtained by taking the limit $m_{f}\to 0$.
\begin{align}
A_{WW}(M^{2}_{W})&=\frac{v^{2}}{768\pi^{2}} \Bigg\{
\bar{g}^{2}\left(59 \bar{g}^{2} - 9\bar{g}^{\prime 2}\right)E +
\frac{1}{3}\left(556\bar{g}^{4} - 75\bar{g}^{2}\bar{g}^{\prime 2} -
3\bar{g}^{\prime 4} + 72\lambda\bar{g}^{2} - 48\lambda^{2}\right)
\nonumber \\&\quad
 + \frac{4\lambda}{\bar{g}^{2}}\left(8\lambda^{2} - 12\lambda
\bar{g}^{2} +
9\bar{g}^{4}\right)\log\frac{M^{2}_{H}}{\mu^{2}}
\nonumber\\&\quad
 + \frac{1}{2\bar{g}^{2}}\left( - 69\bar{g}^{6} -
53\bar{g}^{4}\bar{g}^{\prime 2} + 17\bar{g}^{2}\bar{g}^{\prime 4} +
\bar{g}^{\prime 6}\right)\log\frac{M^{2}_{Z}}{\mu^{2}} \nonumber \\&\quad
 - \frac{1}{2\bar{g}^{2}}\left[49\bar{g}^{6} +
\bar{g}^{4}(72\lambda - 71\bar{g}^{\prime 2}) +
\bar{g}^{2}(17\bar{g}^{\prime 4} - 96\lambda^{2}) + \bar{g}^{\prime 6} +
64\lambda^{3}\right]\log\frac{M^{2}_{W}}{\mu^{2}}
\nonumber\\&\quad
 - 16\left(3\bar{g}^{4} - 4\bar{g}^{2}\lambda +
4\lambda^{2}\right) J_{1}(r_{W}) + \frac{4(99\bar{g}^{6} +
  33\bar{g}^{4}\bar{g}^{\prime 2} - 19\bar{g}^{2}\bar{g}^{\prime 4} -
  \bar{g}^{\prime 6})}{\bar{g}^{2} + \bar{g}^{\prime 2}} J_{1}(r_{WZ})
  \nonumber\\&\quad
\begin{aligned}
    +2\bar{g}^{4} \sum_{\ell=e,\mu,\tau}\bigg\{&
    \left(\frac{3}{4} r_{\ell W} - 2\right)
    \left(E - \log\frac{m^{2}_{\ell}}{\mu^{2}}\right) +
    \frac{r_{\ell W}^{2}}{16} + \frac{1}{2} r_{\ell W} \\&
-\frac{10}{3} +\left(\frac{r_{\ell W}^{3}}{64} -
\frac{3}{4} r_{\ell W} + 2\right)\log\left(1 -
\frac{M^{2}_{W}}{m^{2}_{\ell}}\right)\bigg\}
\end{aligned} \nonumber\\&\quad
\begin{aligned}
    +\frac{8\bar{g}^{2}N_{c}}{v^{2}}\sum_{\alpha,\beta} |K_{\alpha\beta}|^{2} \bigg\{&
\left(3M^{2}_{d_{\beta}} + 3M^{2}_{u_{\alpha}} - 2M^{2}_{W}\right)E \\&
+ \frac{(M^{2}_{d_{\beta}} -
M^{2}_{u_{\alpha}})^{2}}{M^{2}_{W}} + 2(M^{2}_{d_{\beta}} + M^{2}_{u_{\alpha}})-\frac{10}{3}M^{2}_{W}\\&
+\left[\frac{(M^{2}_{d_{\beta}} -
  M^{2}_{u_{\alpha}})^{3}}{2M^{4}_{W}} - \frac{3}{2}(M^{2}_{d_{\beta}} + M^{2}_{u_{\alpha}}) +
M^{2}_{W}\right]\log\frac{M^{2}_{u_{\alpha}}}{\mu^{2}}\\&
+\left[\frac{(M^{2}_{u_{\alpha}} -
  M^{2}_{d_{\beta}})^{3}}{2M^{4}_{W}} - \frac{3}{2}(M^{2}_{d_{\beta}} + M^{2}_{u_{\alpha}}) +
M^{2}_{W}\right]\log\frac{M^{2}_{d_{\beta}}}{\mu^{2}}\\&
+\left[\frac{(M^{2}_{d_{\beta}} - M^{2}_{u_{\alpha}})^{2}}{M^{4}_{W}} +
\frac{(M^{2}_{d_{\beta}} + M^{2}_{u_{\alpha}})}{M^{2}_{W}} - 2\right] J_{3}(M_{u_{\alpha}},M_{d_{\beta}})\bigg\}
\end{aligned}\nonumber \\&\quad
-6\xi_{W}\bar{g}^{4}\left(E + 1 - \log{\xi_{W}} -
\log\frac{M^{2}_{W}}{\mu^{2}}\right) \nonumber \\&\quad
 - 3\xi_{Z}\bar{g}^{2}(\bar{g}^{2} + \bar{g}^{\prime 2})\left(E + 1 -
\log{\xi_{Z}} - \log\frac{M^{2}_{Z}}{\mu^{2}}\right)
\Bigg\} \,,
\end{align}
where 
\begin{equation}
    M_{u}= \operatorname{diag}(m_{u},m_{c},m_{t})\,,\qquad
    M_{d}= \operatorname{diag}(m_{d},m_{s},m_{b})\,,
\end{equation}
$K_{\alpha\beta}$ is the CKM matrix, and the summation indices in the
hadronic contribution run over all the quark generations.
The infinite quantity $E$ is given by eq.~\eqref{eq:e}, and the
functions $J_1(x), J_2(x)$ and $ J_3(x)$ are defined through
\begin{equation}
J_{1}(x)\equiv
\begin{cases}
\frac{\sqrt{1-x}}{x} \log\left(\frac{1+\sqrt{1-x}}{\sqrt{x}}\right)\,,
\quad 0 < x \le 1 \,,\\
-2\frac{\sqrt{x-1}}{x}
\arctan\left(\frac{\sqrt{x-1}}{1+\sqrt{x}}\right)\,, \quad x \ge 1 \,,
\end{cases}
\label{eq:I1}
\end{equation}
\vspace{0.0cm}
\begin{equation}
J_{2}(x)\equiv
\begin{cases}
\sqrt{1-x} \left[ \log\left(\frac{2-x-2\sqrt{1-x}}{x}\right) +i\pi
  \right]\,, \quad 0 < x \le 1 \,,\\
-2\sqrt{x-1}\arctan\left(\frac{1}{\sqrt{x-1}}\right)\,, \quad x \ge 1
\,,
\end{cases}
\label{eq:I2}
\end{equation}
and 
\begin{multline}
J_{3}(M_{u},M_{d})\equiv\sqrt{\left[(M_{d} - M_{u})^{2} -
    M^{2}_{W}\right]\left[(M_{d} + M_{u})^{2} - M^{2}_{W}\right]}\\
\times\log\left[\frac{(M^{2}_{d} + M^{2}_{u} - M^{2}_{W}) +
    \sqrt{\left[(M_{d} - M_{u})^{2} - M^{2}_{W}\right]\left[(M_{d} +
        M_{u})^{2} - M^{2}_{W}\right]}}{2M_{d}M_{u}}\right] \,.
\end{multline}
For completeness we also add here the $W$-boson one-loop self-energy
at zero external momentum, evaluated in Feynman gauge, needed in the
master formula~\eqref{treesmeft}. It reads
\begin{align}
A_{WW}(0) &= \frac{\bar{g}^{4}v^2}{64\pi^2} \Bigg\{
\left(1-\frac{\bar{g}^{\prime 2}}{\bar{g}^{2}}\right)E
+\frac{\lambda}{2 \bar{g}^{2}} - \frac{7 \bar{g}^{\prime 2}}{8
\bar{g}^{2}} + \frac{27}{8}
-\frac{3 \lambda}{(\bar{g}^{2} - 4\lambda)}\log \frac{M^{2}_{h}}{\mu^{2}}
\nonumber\\&\qquad\quad
+ \left( \frac{17
\bar{g}^{2}}{4 \bar{g}^{\prime 2}} + \frac{3 \bar{g}^{2}}{4
(\bar{g}^{2} - 4 \lambda )} - \frac{1}{2}\right) \log
\frac{M^{2}_{W}}{\mu^{2}} 
-\left( \frac{17 \bar{g}^{2}}{4 \bar{g}^{\prime 2}} -
\frac{\bar{g}^{\prime 2}}{\bar{g}^{2}} + \frac{5}{4}\right)
\log\frac{M^{2}_{Z}}{\mu^{2}} 
\Bigg\} \nonumber \\&\quad
\begin{aligned}
+\frac{\bar{g}^{2}N_{c}}{32 \pi^2}\sum_{\alpha,\beta} |K_{\alpha\beta}|^{2}
\Bigg[&\left(M^{2}_{u_{\alpha}} + M^{2}_{d_{\beta}}\right)\left(E -
\log\frac{M^{2}_{d_{\beta}}}{\mu^{2}}\right) \\&
+\frac{M^{2}_{u_{\alpha}}+M^{2}_{d_{\beta}}}{2}
+\frac{M^{4}_{u_{\alpha}}}{M^{2}_{u_{\alpha}} -
M^{2}_{d_{\beta}}}\log\frac{M^{2}_{d_{\beta}}}{M^{2}_{u_{\alpha}}}\Bigg]
\end{aligned}
\nonumber \\&\quad
+\frac{\bar{g}^{2}}{32
\pi^2}\sum_{\ell=e,\mu,\tau}m^{2}_{\ell}\left[
\left(E - \log \frac{m^{2}_{\ell}}{\mu^{2}}\right) + \frac{1}{2}\right]\,.
\label{eq:AWW0}
\end{align}
{Moreover, the derivative of the Higgs self-energy reads}
\begin{align}
\Pi^{\prime}_{HH}(M^{2}_{h})&= \frac{1}{128\pi^{2}}\Bigg\{
(12\bar{g}^2-16\lambda) \left(E-\log\frac{M_{W}^{2}}{\mu^{2}}\right)
+\frac{6}{\lambda}(\bar{g}^{4}+2\bar{g}^{2}\lambda-4\lambda^{2}) \nonumber\\&\quad
+\frac{16\lambda^{3}-20\bar{g}^{2}\lambda^{2}+4\bar{g}^{4}\lambda+3\bar{g}^{6}}
{\lambda\left(\bar{g}^{2} - \lambda\right)}
J_{2}(r_{W}) \nonumber\\&\quad
+\left[6(\bar{g}^{2} + \bar{g}^{\prime 2}) - 8\lambda \right]
  \left(E - \log\frac{M_{Z}^{2}}{\mu^2}\right) +
  \frac{3}{\lambda}\left[(\bar{g}^{2} + \bar{g}^{\prime 2})^{2} +
    2\lambda (\bar{g}^{2} + \bar{g}^{\prime 2}) - 4
\lambda^{2}\right] \nonumber \\&\quad
 + \frac{16\lambda^{3} - 20\lambda^{2}(\bar{g}^{2} +
    \bar{g}^{\prime 2}) + 4\lambda(\bar{g}^{2} + \bar{g}^{\prime
      2})^{2} + 3(\bar{g}^{2} + \bar{g}^{\prime 2})^{3}}
       {2\lambda\left(\bar{g}^{2} + \bar{g}^{\prime 2} -
         \lambda\right)} J_{2}(r_{Z}) + 4\lambda(9 -
         2\sqrt{3}\pi) \nonumber \\&\quad
 - 16\sum_{f}N_{c,f}\left({\frac{m_{f}}{v}}\right)^{2}\left[E -
       \log \frac{m^{2}_{f}}{\mu^{2}} + 1 + r_{f} + \left(1 +
   \frac{r_{f}}{2}\right) J_{2}(r_{f})\right]\nonumber \\&\quad
 + 4\left(4\lambda - \bar{g}^{2}\xi_{W}\right)\left(E -
       \log\frac{M_{W}^{2}}{\mu^{2}} - \log \xi_{W}\right)
       + 4\left(8\lambda - \bar{g}^{2}\xi_{W}\right) +
       16\lambda J_{2}(\xi_{W} r_{W}) \nonumber \\&\quad
 + 2\left[4\lambda - (\bar{g}^{2} + \bar{g}^{\prime
         2})\xi_{Z}\right] \left(E - \log 
       \frac{M_{Z}^{2}}{\mu^{2}} - \log \xi_{Z}\right) 
   + 2\left[8\lambda - (\bar{g}^{2} + \bar{g}^{\prime
       2})\xi_{Z}\right] + 8\lambda J_{2}(\xi_{Z} r_{Z})\Bigg\} \,,
\end{align}
and the light quark contribution needed in \eq{eq:pgghad} is
\begin{equation}
\label{app:Pggh}
\Pi_{\gamma\gamma}^{\text{had}}(M_{Z}^{2})= \frac{\bar{g}^{2}g^{\prime
2}}{12\pi^{2}(\bar{g}^{2}+\bar{g}^{\prime 2})} \sum_{q}N_{c}Q_{q}^{2}\left[
E-\log\frac{m_{q}^{2}}{\mu^{2}}
+\left(1+\frac{r_{qZ}}{2}\right)J_{2}(r_{qZ})+ r_{qZ} +\frac{5}{3} \right] \,.
\end{equation}

\bibliography{EFT}{}
\bibliographystyle{JHEP}

\end{document}